 \documentclass[final,1p,times]{elsarticle}

\usepackage[utf8]{inputenc}
\usepackage{graphicx}
\usepackage{amssymb}
\usepackage{pifont}
\usepackage{subcaption}
\usepackage{booktabs}
\usepackage{threeparttable}
\usepackage[hidelinks]{hyperref}

\usepackage[svgnames]{xcolor}

\usepackage{lineno}

\biboptions{comma,square,sort}
\journal{Big Data Research}

\begin{document}

\begin{frontmatter}


\title{Theodolite: Scalability Benchmarking of Distributed Stream Processing Engines in Microservice Architectures}



\author[se]{Sören Henning\corref{cor1}}
\ead{soeren.henning@email.uni-kiel.de}

\author[se]{Wilhelm Hasselbring}
\ead{hasselbring@email.uni-kiel.de}

\address[se]{Software Engineering Group, Kiel University, 24098 Kiel, Germany}

\cortext[cor1]{Corresponding author}

\begin{abstract}
Distributed stream processing engines are designed with a focus on scalability to process big data volumes in a continuous manner.
We present the Theodolite method for benchmarking the scalability of distributed stream processing engines. Core of this method is the definition of use cases that microservices implementing stream processing have to fulfill. For each use case, our method identifies relevant workload dimensions that might affect the scalability of a use case. We propose to design one benchmark per use case and relevant workload dimension.

We present a general benchmarking framework, which can be applied to execute the individual benchmarks for a given use case and workload dimension. Our framework executes an implementation of the use case's dataflow architecture for different workloads of the given dimension and various numbers of processing instances. This way, it identifies how resources demand evolves with increasing workloads.

Within the scope of this paper, we present 4 identified use cases, derived from processing Industrial Internet of Things data, and 7 corresponding workload dimensions. We provide implementations of 4 benchmarks with Kafka Streams and Apache Flink as well as an implementation of our benchmarking framework to execute scalability benchmarks in cloud environments. We use both for evaluating the Theodolite method and for benchmarking Kafka Streams' and Flink's scalability for different deployment options.
\end{abstract}

\begin{keyword}
Stream Processing \sep Microservices \sep Benchmarking \sep Scalability


\end{keyword}

\end{frontmatter}


\section{Introduction}

The era of big data with its immense volume of data and often varying or unpredictable workloads requires software systems to ``scale out'', for example, by being distributed among multiple computing nodes in elastic cloud environments \cite{Hashem2015}.
In order to make software systems scalable, software architects apply design patterns such as microservices and event-driven architectures \cite{Katsifodimos2019, Hasselbring2016}.
In such architectures, loosely coupled components, which are separated by business functions (microservices), often communicate with each other primarily asynchronously via a dedicated messaging system \cite{Hasselbring2017}.
Within individual microservices, incoming data has to be processed by transforming, aggregating, and joining with other data. Often the results of such operations are again published to the messaging system, allowing other services to subscribe to these data. To process data within microservices, stream processing engines such as Apache Samza \cite{Noghabi2017} or Apache Kafka Streams \cite{Sax2018} are increasingly used. With such tools, data is processed in operators, which are connected to form directed acyclic graphs, describing the dataflow within services.
Scalability of these microservices is then achieved by letting the individual instances of a microservice process only a part of the data (data parallelism) or execute only a part of the operations (task parallelism) \cite{Roger2019}.

However, a huge challenge when building and evaluating such big data architectures is to determine how they will scale with increasing workload.
The scalability of a microservice that applies stream processing techniques might depend on the selected stream processing engine, but also on the choice of deployment options, for example, concerning the engine's configuration, the messaging system, and the cloud environment.
The multitude of deployment options makes evaluating and fine-tuning such microservices with different implementations and deployment options expensive in time and resources \cite{Akidau2019}.
In empirical software engineering \cite{Tichy1998}, often benchmarks are used as a measuring instrument for comparing different technologies or configurations \cite{Sim2003, Bermbach2017a}.
Combined with search-based software engineering, benchmarks allow making and evaluating decisions regarding a software's architecture, its implementation, and its deployment \cite{Frey2013, Bauer2019}.

Whereas benchmarking performance qualities of stream processing engines such as throughput or latency is heavily performed by academia and industry \cite{Chintapalli2016, Karimov2018, VanDongen2020}, approaches on benchmarking their scalability do not exist so far.
With this paper, we make the following contributions:

\begin{enumerate}
	\item We present Theodolite,\footnote{A theodolite is an optical instrument used in geodesy for measuring angles.}
	the first method for benchmarking the scalability of stream processing engines.
	\item As this method aims to create specification-based benchmarks \cite{Kistowski2015}, we identify common use cases for stream processing within microservices, which are inspired by real industrial settings in the context of our Titan project \cite{Hasselbring2019}.
	\item We argue that for generating a benchmark's workload, different dimensions of workloads should be considered. For our identified use cases, we identify different workload dimensions, a stream processing engine may or may not scale with.
	\item We propose a benchmarking framework, which can be used to benchmark a system under test (SUT) for a selected use case and a selected workload dimension.
	\item We provide benchmark implementations for all identified use cases with Kafka Streams and Apache Flink as well as an implementation of our benchmarking framework as open source.\footnote{\url{https://github.com/cau-se/theodolite}}
	\item We exemplify the use of these benchmark implementations for evaluating the impact of different deployment options for Kafka Streams and Flink applications on scalability.
	A replication package and the collected data of our experiments is published as supplemental material \cite{ReplicationPackage}, such that other researchers may repeat and extend our work.
\end{enumerate}

The remainder of this paper is structured as follows. Section~\ref{sec:Related} starts by placing this paper in the context of related work.
Section~\ref{sec:method} describes our proposed benchmarking method.
Section~\ref{sec:UseCases} identifies use cases for stream processing and Section~\ref{sec:Dimensions} identifies relevant workload dimensions for these use cases.
Section~\ref{sec:metrics} discusses our proposed scalability metrics and measurement methods.
Section~\ref{sec:framework} describes our benchmarking framework architecture, which can be applied to benchmark a given use case with a given workload dimension.
Section~\ref{sec:implementation} presents our implementation of this architecture including concrete benchmark implementations,
used for our evaluation in Section~\ref{sec:Evaluation}.
Finally, Section~\ref{sec:Conclusions} concludes this paper and points out future work.

\section{Related Work}\label{sec:Related}

Big data stream processing, scalability in cloud computing, and benchmarking big data systems as well as cloud services are active fields of research. Subject of this paper is the intersection of the research fields on stream processing, scalability, and benchmarking. To the best of our knowledge, we present the first work on benchmarking scalability of stream processing engines. In the following, we relate our paper to work on scalable stream processing, benchmarking stream processing engines, and benchmarking scalability.

\subsection{Scalable Stream Processing}\label{sec:stream-processing-foundations}

Closely related to the emergence of real-time stream processing systems are requirements for scalability \cite{Stonebraker2005}.
Most modern stream processing engines apply dataflow models \cite{Akidau2015, Sax2018} that adopt the MapReduce \cite{DeanGhemawat2010} approach to continuous data streams. That is, their primary concept for achieving scalability is to require each message to contain a key, which is used to partition the data stream. Thus, individual stream partitions can be processed in parallel.
Individual instances of stream processing operators only receive data for a certain set of keys, allowing as many operator instances to be deployed as there are different keys.
An additional factor of scaling can be obtained by executing multiple stream processing operators in parallel or with multiple elasticity levels \cite{ESOCC2015}. \citet{Roger2019} present a comprehensive survey on parallelization approaches in stream processing.

In the context of stream processing, some scalability evaluations were performed, for example, when presenting new stream processing engines \cite{Akidau2013, Kulkarni2015}, streaming operators \cite{Dossinger2019, Karimov2019, Benson2020}, or stream processing architectures \cite{Henning2019a, Henning2020}. However, these evaluations do not present a systematic approach to benchmark entire or arbitrary stream processing engines.

\subsection{Benchmarking Stream Processing Engines}


Several studies have been conducted that benchmark performance metrics such as throughput and latency \cite{Lu2014, Chintapalli2016, Shukla2017, Lopez2016, Karakaya2017, Karimov2018, Hesse2018, Nasiri2019, Reichelt2019, Shahverdi2019, Bordin2020, Pagliari2020, VanDongen2020, Vikash2020} of stream processing engines. Many studies focus on benchmarking single streaming operators, often including data ingestion and results publishing, in micro-benchmarks \cite{Lu2014, Shukla2017, Karimov2018, Reichelt2019} or construct benchmarks based on technical capabilities of stream processing engines \cite{Chintapalli2016, Karakaya2017, VanDongen2020}. In this paper, we create benchmarks based on use cases for stream processing within microservices. Our use cases process Industrial Internet of Things (IIoT) data, similar to \citet{Hesse2018}. Internet of Things (IoT) data, for example, smart city and smart health data, is also used for performance benchmarks \cite{Shukla2017, Nasiri2019, VanDongen2020, Vikash2020}.

The objective of most benchmarking studies is to compare different stream processing engines.
Additionally, some studies include experiments of different deployment options (e.g., release versions \cite{Lu2014, Chintapalli2016}, processing guarantees \cite{Chintapalli2016, Nasiri2019}, or messaging system configurations \cite{Karakaya2017}).
In this paper, we focus on benchmarking different deployment options (see our experimental evaluation in Section~\ref{sec:Evaluation}), for example, to find optimal ones as it is done by \citet{Frey2013}.

Most studies benchmark the stream processing engines Apache Storm \cite{Toshniwal2014}, Apache Spark \cite{Zaharia2013}, and Apache Flink \cite{Carbone2015}.
Besides Apache Flink, we select the rather new Apache Kafka Streams for our experimental evaluation as it is explicitly designed for stream processing within microservices. Benchmarks of Kafka Streams were only performed in two recent studies \cite{Shahverdi2019, VanDongen2020}.
Whereas some studies \cite{Lu2014, Chintapalli2016, Shukla2017} benchmark stream processing engines with their default configurations, others highly optimize their configurations for their specific test scenario \cite{Karimov2018, VanDongen2020} or even per workload \cite{Karakaya2017}. As we benchmark different deployment options in our experimental evaluations, we mainly stay with default configurations, except the ones that should be optimized.

Most benchmark setups \cite{Lu2014, Chintapalli2016, Karakaya2017, Lopez2016, Hesse2018, Nasiri2019, Shahverdi2019, Bordin2020, Pagliari2020, VanDongen2020} include a messaging system, namely Apache Kafka \cite{Kreps2011}, as a middleware component between workload generation and stream processing engine. \citet{Karimov2018} leave out such a system as they argue that it may become the bottleneck of the benchmark, whereas we consider it necessary in a use-case-oriented benchmark design.
Only one recent benchmarking study \cite{VanDongen2020} executes benchmarks fully containerized in a cloud environment instead of virtual machines or bare-metal servers. Our benchmark implementations presented in Section~\ref{sec:implementation} are also intended to be deployed fully containerized in a cloud environment since such a deployment matches most closely future execution environments.

The fact that all modern stream processing engines are designed for execution in a distributed fashion is respected in most benchmarks. Most benchmarking studies execute multiple instances of the stream processing engine, distributed in a cluster of multiple nodes.
Further, some studies \cite{Karimov2018, Lopez2016, Karakaya2017, Nasiri2019, Vikash2020} evaluate how performance evolves when scaling out the stream processing cluster (i.e., increasing the amount of processing instances).
For this purpose, \citet{Karimov2018} introduce the definition of sustainable throughput, which is the maximum throughput a stream processing engine can process without discarding or queuing up data over a longer period of time. The authors determine the sustainable throughput for each evaluated cluster size and generate data according to that throughput. Our proposed scalability measurement method in Section~\ref{sec:measuring-scalability} adopts a concept similar to sustainable throughput, but with the difference that we determine the required processing resources for fixed workloads.
\citet{Lopez2016} explicitly investigate how throughput evolves with different numbers of CPU cores.
\citet{Karakaya2017} and \citet{Vikash2020} present the ``scale-up ratio'' metric, giving, for a number of processing instances, the percentage increase of processed elements compared to processing with one instance.
Worth mentioning is also the work of \citet{Zeuch2019}, who evaluate whether scaling up stream processing deployments can be an alternative to scaling out. For this purpose, they compare the performance of stream processing engines in single-node deployments on modern hardware.
Again, all of these studies evaluate scalability in function of available resources and not in function of different workload scales, as we suggest in this paper due to common scalability definitions. 

Some studies \cite{Lu2014, Chintapalli2016, Karakaya2017, Nasiri2019, Shahverdi2019} also evaluate how performance evolves when scaling the workload.
However, these studies do not determine the resource demand for the individual tested workloads. They do not relate workload and resource requirements and, therefore, cannot provide a measure to assess scalability.
Even though there exists no scalability benchmark for stream processing engines so far, the requirement for benchmarking their scalability was already identified in some studies \cite{VanDongen2020, Lu2014}.

\subsection{Benchmarking Scalability}

Although not for stream processing, benchmarking scalability has already been addressed for other types of big data software.
\citet{Rabl2012} designed benchmarks for distributed database systems, which were, inter alia, used for a thorough analysis of the tested systems' scalability. Such distributed database systems are highly related to distributed stream processing engines as both apply similar concepts for data partitioning. Also, they are often used in conjunction in big data software systems to materialize processed streaming data \cite{Ranjan2014}. Later, \citet{Kuhlenkamp2014} reproduced and extended these works by explicitly benchmarking scalability and elasticity. Besides comparing different database systems, the authors also benchmarked different deployment options, like we do in this paper.

A precise differentiation between scalability and elasticity is given by \citet{Herbst2013}. Several scalability metrics and measurement methods have been proposed \cite{Lehrig2015, Sanders2015, Brataas2017, Bermbach2017a}. Similar to the \textit{scalability rate} defined by \citet{Sanders2015}, our proposed metric for scalability in stream processing (see Section~\ref{sec:metrics}) is also a function. However, instead of a function of available resource, we consider it as a function of the workload (see Section~\ref{sec:measuring-scalability}). That is, we evaluate how the number of processing instances evolves with increasing workload.
In the context of benchmarking the scalability of cloud services, it is often also analyzed how costs evolve with increasing workload \cite{Sanders2015, Brataas2017}.

\section{Benchmarking Method}\label{sec:method}

In this section, we present our Theodolite method for benchmarking scalability of stream processing engines in microservices.

\subsection{Use-Case-oriented Benchmark Design}\label{sec:benchmark design}
Microservices are usually the smallest deployable entity and can only be scaled at a whole. We therefore focus on application-driven benchmarks \cite{Bermbach2017a} (also referred to as macro-benchmarks) 
that evaluate the scalability of an entire microservice instead of individual processing steps.
A key requirement for benchmarks is relevance \cite{Kistowski2015}. Therefore, a benchmark should represent a typical use case of stream processing within microservices which is required to be scalable.
Furthermore, our benchmarking approach is specification-based \cite{Kistowski2015}, meaning that use cases are defined based on functional or business requirements instead of technical ones.
We identify four common use cases in Section~\ref{sec:UseCases}.
As a messaging system is an integral part of most event-driven big data architectures, we also include the messaging system in all our benchmarks.

\subsection{Distinction between Workload Dimensions}\label{sec:distinct-dimensions}

Scalability is usually defined as the ability of a system to continue processing an increasing workload with additional resources provided \cite{Herbst2013}. However, most definitions do not specify the term ``load'' more precisely \cite{Herbst2013, Lehrig2015, Bermbach2017a}.
In fact, load can be increased or decreased in various dimensions \cite{Duboc2007}.
For example, one thousand data sources generating one message per second would generate an overall load on a stream processing engine of one thousand messages per second. Scaling up the amount of data sources to one million or scaling up the message frequency to one message per millisecond would both lead to an overall load of one million messages per second. However, it is way more likely that a stream processing engine, which aggregates messages per data source, scales better with the amount of data sources.\footnote{provided that input messages are keyed by their data source, which is a natural choice in data streaming}
Thus, it is necessary to explicitly define for which load dimension a benchmark should test scalability.
Realistic workloads may be based on observing software systems in production \cite{WESSBAS2018}.

We present a list of different workload dimensions in Section~\ref{sec:Dimensions}. There, it also becomes apparent that the set of applicable workload dimensions always depends on the use case. We therefore suggest to design one benchmark per use case and workload dimension. Nevertheless, our proposed scalability metrics and measurement methods can be applied to all benchmarks.
In Section~\ref{sec:framework}, we therefore present a benchmarking framework architecture that can be configured with a use case and a workload dimension.

\subsection{Measuring Scalability}\label{sec:measuring-scalability}

Following common scalability definitions, benchmarking scalability has to assess whether a SUT is able to process an increasing amount of data with additionally provided resources. Thus, our benchmarking method evaluates how the resource demand evolves with increasing workloads. As we study horizontal scalability \cite{Abbott2015} in this paper, increasing available resources corresponds to providing additional processing instances.

\begin{figure}[t]
	\centering%
	\includegraphics[width=0.4\linewidth]{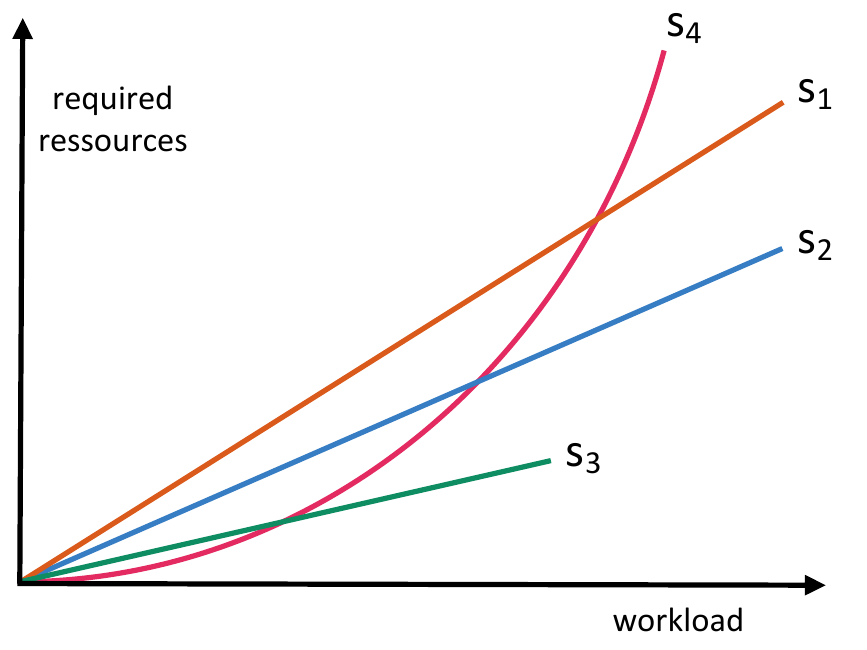}%
	\caption{Example of a scalability graph for four systems.}
	\label{fig:example-scalability-graph}
\end{figure}%

The basic concept of our benchmark design is to generate different workloads and to experimentally determine the required number of processing instances for each workload. Based on these experiments, a scalability graph (see Figure~\ref{fig:example-scalability-graph})
can be obtained, which shows how the resource demand evolves with an increasing load.
Respecting statistical requirements such as a sufficiently large sample size, conclusions can be drawn about the scalability behavior of the SUT. For example, the scalability graph allows to conclude whether it scales linearly (S$_1$, S$_2$, and S$_3$ in Figure~\ref{fig:example-scalability-graph}), quadratically (S$_4$ in Figure~\ref{fig:example-scalability-graph}), etc. and with which factors.
Moreover, the scalability graph can be used to identify a critical point, beyond which the system does not scale (S$_3$ in Figure~\ref{fig:example-scalability-graph}). Our proposed method to determine the number of required instances per workload is presented in Section~\ref{sec:metrics}.

\subsection{Systems under Test (SUT)}

The systems under test (SUT) we consider in this paper are microservices that perform stream processing. Thus, we characterize a SUT by the stream processing engine used for implementation as well as a set of selected deployment options. 

Comparing different stream processing engines is probably the most common type of benchmarking studies for big data applications (see Section~\ref{sec:Related}). Such benchmarks may serve for deciding which stream processing engine to use in a project.
However, a particular stream processing engine is often already chosen based on other criteria, for example, the API's ease of use, community size, availability of developers, or commercial support. Instead, it can be more important to choose appropriate deployment options for the employed stream processing engine.

Modern stream processing engines provide a plethora of configuration options. For example, they allow to configure different buffer sizes and commit rates. Finding a good configuration can thus be a tedious task.
Benchmarking different configurations against each other can assist in optimizing the configuration, without running entire applications.
Similarly, benchmarking scalability of different execution environments helps to find a suitable execution environment. For example, benchmarking different cloud providers allows to assess with which provider an application may scale best \cite{Bermbach2017b}. This also applies for evaluating different configurations of infrastructure services such as a messaging system.

\section{Identification of Use Cases}\label{sec:UseCases}

\begin{figure}[t]
	\centering%
	\includegraphics[width=\linewidth]{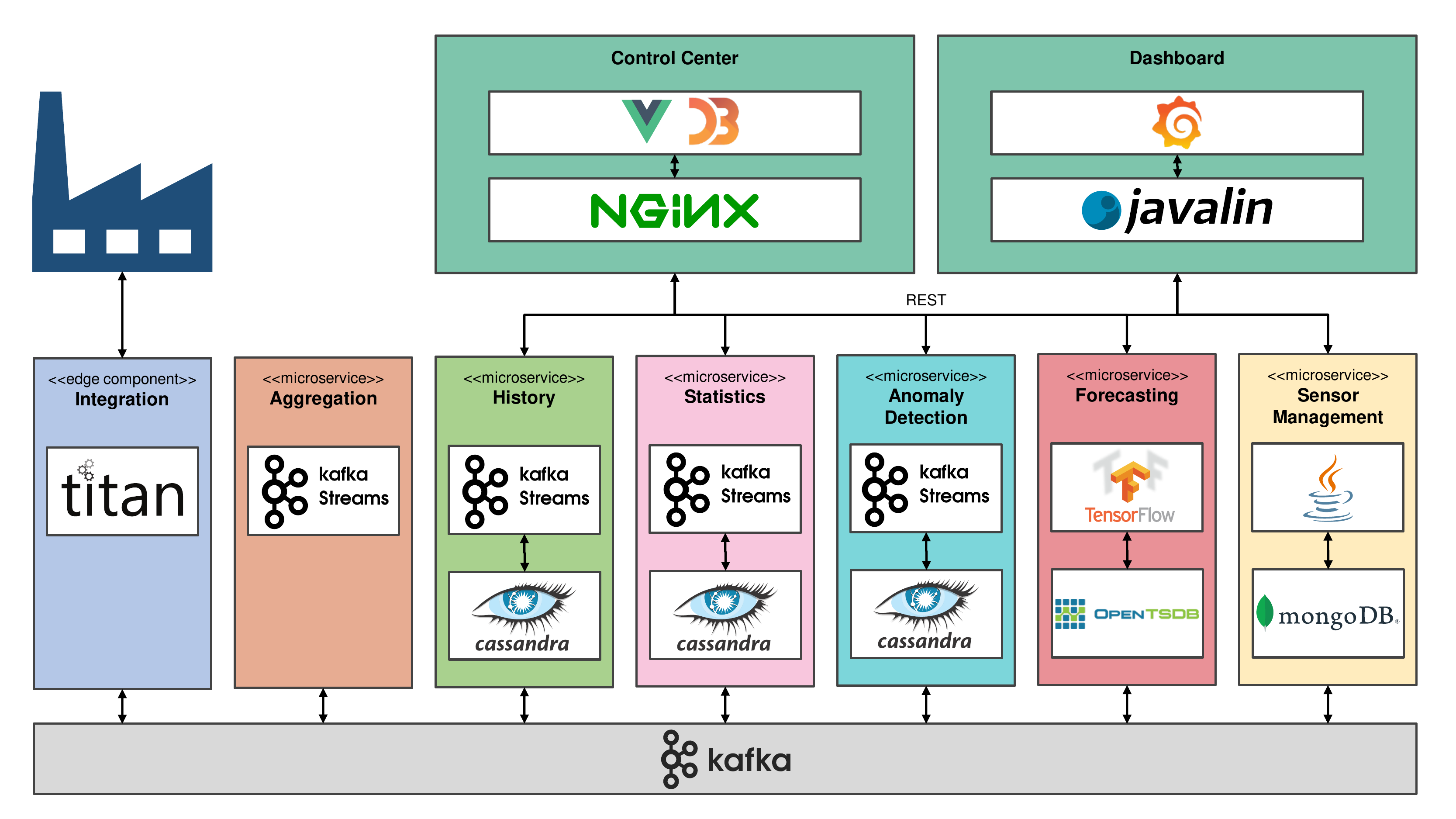}%
	\caption{Microservice architecture of the Titan Control Center serving as a reference of our identified use cases \cite{ArxivGoalsMeasures}.}
	\label{fig:titan-cc-architecture}
\end{figure}

In this section, we identify four use cases of different complexity for stream processing engines deployed as microservices. 
Our use cases are derived from the Titan Control Center \cite{Henning2019a, Henning2021, ArxivGoalsMeasures}, a microservice-based analytics platform that performs different kinds of analyses on Industrial Internet of Things (IIoT) data (see Figure~\ref{fig:titan-cc-architecture}).
Although derived from IIoT, we assume that these use cases also occur in other application domains.
For each use case, we present a corresponding dataflow architecture (often referred to as operator graph or topology). Our presented architectures do not follow a specific model as different stream processing engines use different models \cite{Akidau2015, Sax2018}.
However, most of these models are similar, allowing the described architectures to be implemented in most modern engines.

All our use cases share that they receive all data from a messaging system and publish all processing results back to that messaging system.
We assume all input messages to be measurements from IIoT sensors. They are keyed by an identifier of their corresponding sensor and contain the actual measurement as value.
The dataflow architectures presented below focus only on required processing steps. In practice, microservices fulfilling these use cases are likely to contain additional processing steps, for example, for filtering and transforming intermediate data.
Table~\ref{tab:dataflow-characteristics} lists typical dataflow characteristics in stream processing and shows in which use cases these characteristics apply.

\begin{table}
	\centering
	\begin{threeparttable}
		\caption{Overview of dataflow characteristics observed in use cases.}
		\label{tab:dataflow-characteristics}
		\small
		\newcommand{\cmark}{\ding{51}}%
		\newcommand{\xmark}{\ding{55}}%
		\begin{tabular}{lcccc}
			\toprule 
			Dataflow characteristics & UC1 & UC2 & UC3 & UC4 \\
			\midrule 
			Stateless operations & \cmark & \cmark & \cmark & \cmark \\
			Tumbling window aggregations & \xmark & \cmark & \xmark & \cmark\tnote{a} \\
			Sliding window aggregations & \xmark & \xmark & \cmark & \cmark\tnote{a} \\
			Joins of different streams & \xmark & \xmark & \xmark & \cmark \\
			Feedback loops & \xmark & \xmark & \xmark & \cmark \\
			\bottomrule
		\end{tabular}
		\begin{tablenotes}\footnotesize
			\item [a] Use case UC4 can be configured either to use tumbling windows or sliding windows. 
		\end{tablenotes}
	\end{threeparttable}
\end{table}

\subsection{Use Case UC1: Database Storage}

A simple, but common use case in event-driven architectures is that events or messages should be stored permanently, for example, in a NoSQL database. Using this database, an application can provide its data via an API as it is the case in Lambda and Kappa architectures \cite{Lin2017}. In the Titan Control Center, for example, the Statistics and the History microservices store processed data this way (see Figure~\ref{fig:titan-cc-architecture}).

\begin{figure}[t]
	\centering%
	\includegraphics[width=0.5\linewidth]{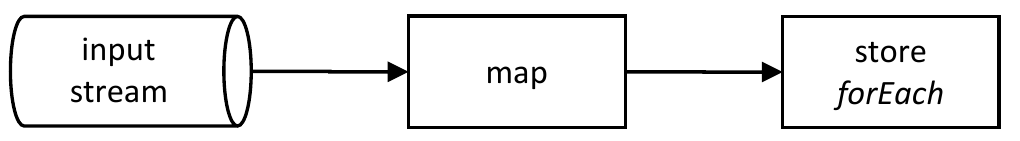}%
	\caption{Dataflow architecture for UC1: Database Storage.}
	\label{fig:architecture-uc1}
\end{figure}%

A dataflow architecture for this use case is depicted in Figure~\ref{fig:architecture-uc1}. The first step is to read data records from a messaging system. Then, these records are converted into another data format in order to match the often different formats required by the database. Finally, the converted records are written to an external database. Depending on the required processing guarantees, this is done either synchronous or asynchronous. Unless the interaction between database and stream processing engine should be benchmarked, we suggest to not include a real database in implementations of these benchmarks. Otherwise, due to the simple, stateless stream processing topology, the benchmark would primarily test the database's write capabilities.

\subsection{Use Case UC2: Downsampling}

A very common use case for stream processing architectures is reducing the amount of events, messages, or measurements by aggregating multiple records within consecutive, non-overlapping time windows.
Typical aggregations compute the average, minimum, or maximum of measurements within a time window or count the occurrence of same events. Such reduced amounts of data are required, for example, to save computing resources or to provide a better user experience (e.g., for data visualizations). When using aggregation windows of fixed size that succeed each other without gaps (called tumbling windows \cite{Carbone2019}), the (potentially varying) message frequency is reduced to a constant value. 
This is also referred to as downsampling. Downsampling allows for applying many machine learning methods that require data of a fixed frequency.
In the Titan Control Center, the History microservice continuously downsamples IIoT monitoring data and provides these data for other microservices (see Figure~\ref{fig:titan-cc-architecture}).

This use case is the basic application for approaches on windowed aggregation.
Since in this work we consider use cases from a functional requirements perspective rather than from a technical one, we do not equate this use case with windowed aggregations. Indeed, windowed aggregations might even be used in other use cases (see UC3 and UC4).

\begin{figure*}[t]
	\centering%
	\includegraphics[width=0.75\textwidth]{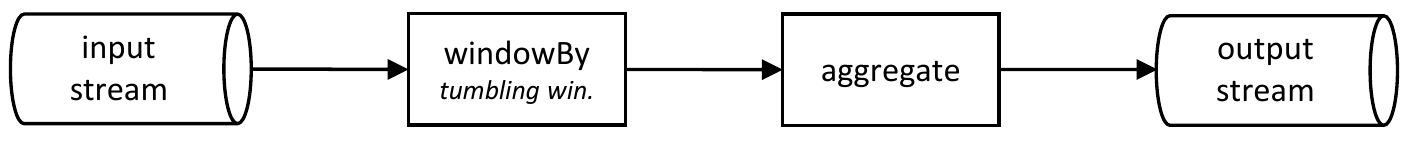}%
	\caption{Dataflow architecture for UC2: Downsampling}
	\label{fig:architecture-uc3}
\end{figure*}%

A dataflow architecture for this use case is presented in Figure~\ref{fig:architecture-uc3}. It first reads measurement data from an input stream and then assigns each measurement to a time window of fixed, but statically configurable size. Afterwards, an aggregation operator computes the summary statistics sum, count, minimum, maximum, average, and population variance for a time window. Finally, the aggregation result containing all summary statistics is written to an output stream.

\subsection{Use Case UC3: Aggregation based on Time Attributes}
A second type of temporal aggregation is aggregating messages that have the same time attribute. Such a time attribute is, for example, the hour of day, day of week, or day in the year. This type of aggregation can be used to compute, for example, an average course over the day, the week, or the year. It allows to demonstrate or discover seasonal patterns in the data. The Statistics microservice of the Titan Control Center implements this use case (see Figure~\ref{fig:titan-cc-architecture}).

This use case differs from UC2 in that the time attribute has to be extracted from the record's timestamp, whereas in UC2 the timestamp needs no further interpretation. Moreover, in this use case, multiple aggregations have to be performed in parallel (e.g., maintain intermediate results for all 7 days of the week). Thus, the amount of different output keys increases by the factor of possible different time attributes. For example, when computing aggregations based on the day of week for a data stream with $n$ different keys, the result stream contains data of $7n$ different keys.

In practice, not all messages that have ever been recorded should be considered in the aggregation, but usually only those of a certain past time period. For example, in industrial facilities, operators are interested in the average course of energy consumption over the day within the last 4 weeks. They do probably not want to include older data as the average course might change over time, for example, due to changing process planing and varying load over the year. Therefore, the aggregation based on time attributes is performed on a sliding window \cite{Carbone2019}, which only considers data of a fixed time period.
In contrast to the tumbling window aggregation in UC2, this use case additionally requires to compute results for multiple overlapping time windows, which further increases the amount of output data.

\begin{figure*}[t]
	\centering%
	\includegraphics[width=\textwidth]{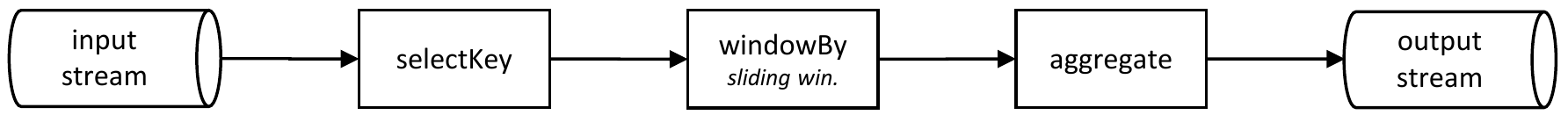}%
	\caption{Dataflow architecture for UC3: Aggregation based on Time Attributes}
	\label{fig:architecture-uc4}
\end{figure*}%

Figure~\ref{fig:architecture-uc4} depicts a dataflow architecture for this use case. The first step is to read measurement data from the input stream. Then, a new key is set for each message, which consists of the original key (i.e., the identifier of a sensor) and the selected time attribute (e.g., day of week) extracted from the record's timestamp. In the next step, the message is duplicated for each sliding window it is contained in. Then, all measurements of the same sensor and the same time attribute are aggregated for each sliding time window by computing the summary statistics sum, count, minimum, maximum, average, and population variance. The aggregation results per identifier, time attribute, and window are written to an output stream.

\subsection{Use Case UC4: Hierarchical Aggregation}

For analyzing sensor data, often not only the individual measurements of sensors are of interest, but also aggregated data for groups of sensors.
When monitoring energy consumption in industrial facilities, for example, comparing the total consumption of machine types often provides better insights than comparing the consumption of all individual machines.
Additionally, it may be necessary to combine groups further into larger groups and adjust these group hierarchies at runtime. A detailed description of these requirements, supplemented with examples, is provided in a previous paper \cite{Henning2019b}.
In the Titan Control Center, the Aggregation microservice hierarchically aggregates IIoT data this way (see Figure~\ref{fig:titan-cc-architecture}).

\begin{figure*}[t]
	\centering%
	\includegraphics[width=\textwidth]{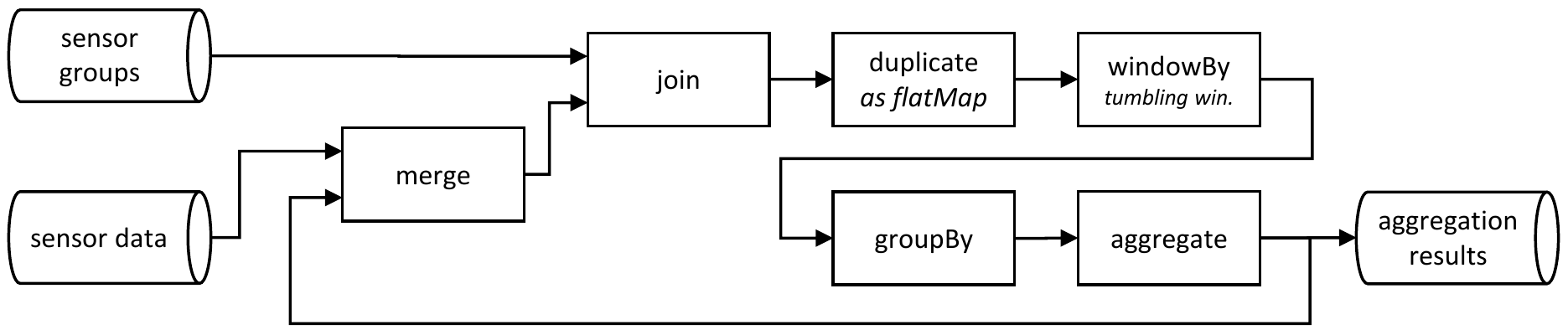}%
	\caption{Dataflow architecture for UC4: Hierarchical Aggregation. This illustration is based on a previous paper \cite{Henning2020}, which also provides a comprehensive description of this architecture.}
	\label{fig:architecture-uc2}
\end{figure*}%

In previous work \cite{Henning2020}, we presented a dataflow architecture for the use case of hierarchically aggregating data streams (see Figure~\ref{fig:architecture-uc2}).
The dataflow architecture requires two input data streams: a stream of sensor measurements and a stream tracking changes to the hierarchies of sensor groups. In the consecutive steps, both streams are joined, measurements are duplicated for each relevant group, assigned to time windows, and the measurements for all sensors in a group per window are aggregated. Finally, the aggregation results are exposed via a new data stream. Additionally, the output stream is fed back as an input stream in order to compute aggregations for groups containing subgroups.
To also support unknown record frequencies, this dataflow architecture can be configured to use sliding windows instead of tumbling windows \cite{Henning2020}.

\section{Identification of Workload Dimensions}\label{sec:Dimensions}

A software system can be considered scalable if it is able to handle an increasing load with additional provided resources. In this section, we give tangible form to the term ``load'' in the context of stream processing. We show that, depending on the particular use case, an increasing load can have different dimensions with which a stream processing application potentially scales differently. In the following, we describe the most important workload dimensions for the use cases in Section~\ref{sec:UseCases}. Table~\ref{tab:benchmarks} relates the individual workload dimensions to the use cases whose scalability they could potentially affect.
Note that the set of presented workload dimensions is not intended to be exhaustive. In particular if further use cases are considered, it is likely that load can increase in further dimensions. Moreover, it may also be required to consider scalability according to multiple workload dimensions \cite{Duboc2007}. In such cases, however, it might be advisable to introduce a cost function to obtain a two-dimensional scalability graph.

\begin{table}
	\centering
	\caption{Overview of workload dimension impacting the individual use case.}
	\label{tab:benchmarks}
	\small
	\newcommand{\cmark}{\ding{51}}%
	\newcommand{\xmark}{\ding{55}}%
	\begin{tabular}{lcccc}
		\toprule 
		Workload Dimension & UC1 & UC2 & UC3 & UC4 \\
		\midrule 
		Message Frequency & \cmark & \cmark & \cmark & \cmark  \\
		Amount of Different Keys & \cmark & \cmark & \cmark & \cmark \\
		Time Window Size & \xmark & \cmark & \cmark & \cmark \\
		Amount of Overlapping Windows & \xmark & \xmark & \cmark & (\cmark) \\
		Amount of Time Attribute Values & \xmark & \xmark & \cmark & \xmark \\
		Number of Elements in Groups & \xmark & \xmark & \xmark & \cmark \\
		Number of Nested Groups & \xmark & \xmark & \xmark & \cmark \\
		\bottomrule
	\end{tabular}
\end{table}

\subsection{Message Frequency}
The message frequency describes the number of messages sent per key (sensors in our use cases) and time. It is the inverse to the time period between messages of the same key. This workload dimension applies to all use cases.

\subsection{Amount of Different Keys}
Another typical benchmark objective is to evaluate how an application scales when increasing the number of keys. In our use cases, this applies if the amount of data sources (e.g., sensors) increases. As all presented use cases require data to be keyed by the sensor identifier, this dimension applies to all of them.
For use cases with stateful data processing (UC2, UC3, and UC4), also the overall state size increases with increasing number of keys, which may impact the overall scalability.

\subsection{Time Window Size}
For all use cases that perform time-related operations based on time windows (UC2, UC3, and UC4), a highly relevant workload dimension is the window size. A larger window size typically means less (intermediate) results and, thus, less transmitted data but also later and perhaps less relevant results.
Whether time window sizes impact scalability probably depends on whether a stream processing engine employs incremental or cumulative techniques for partial aggregations \cite{Carbone2019}.

\subsection{Amount of Overlapping Windows}
Another workload dimension for use cases performing operations on sliding windows (UC3 and optionally UC4) is the amount of overlapping windows. The number of overlapping windows corresponds to the ratio of window size and advance period.
The more overlapping windows are maintained, the more computations for a single message have to be performed.
For UC3 this means that if measurements are aggregated to obtain an average value per hour of day for four weeks, the load on the respective application depends on whether a new computation should be started, for example, every day or only every week. In these cases, the amount of overlapping windows would be either 28 or only 4.
Furthermore, with more overlapping windows, also the state per key and, thus, the overall state size might increase. The influence on scalability is likely to depend on whether a stream processing engine employs window slicing techniques \cite{Carbone2019}.

\subsection{Amount of Time Attribute Values}
The load on a generic approach to aggregate records based on time attributes, as presented in UC3, might depend on the choice of the time attribute. If, for example, the day of week is the selected time attribute, the corresponding architecture would produce 7 results per measurement key. For the time attributes hour of day or day in the year, 24 or 365 results would be produced, respectively. Even larger amounts of time attribute values are conceivable if multiple attributes are combined to a new time attribute, for example, hour of the week to compute a weekly course of hourly resolution.
A benchmark objective could be to evaluate how an application scales with the number of possible values for a time attribute.
While more time attribute values do not produce more outputs, they still have an influence on the state size per key.

\subsection{Maximal Number of Elements in Groups}
The dataflow architecture for the hierarchical aggregation of UC4 only aggregates measurements of those sensors contained in the provided hierarchy. Therefore, it is more likely that only the amount of sensors contained in the hierarchies affects scalability. One parameter controlling the size of the actually aggregated sensors is, thus, the number of elements (sensors or subgroups) in a group.
Hence, this workload dimension does also influence the overall state size.
As groups may have different sizes, we focus on the maximal number of elements in a group.

\subsection{Maximal Depth of Nested Groups}
A second parameter controlling the hierarchy sizes is the number of nested groups.
It also impacts the overall state size.
As again not all branches of the hierarchy necessarily have the same depths, we focus on the maximal depth within hierarchies.

\section{Metrics and Measurement Methods}\label{sec:metrics}

In Section~\ref{sec:measuring-scalability}, we argue that for benchmarking scalability, it is necessary to determine how resource demand increases with increasing workload. We propose a scalability graph (Figure~\ref{fig:example-scalability-graph}) showing the required number of instances for each tested workload. In this section, we define our scalability metric \textit{required number of instances per workload} and present a corresponding measurement method. This method requires additional auxiliary metrics, which we also define in this section. For each auxiliary metrics, we present corresponding measurement methods.

\paragraph{Scalability Metric: Required Number of Instances per Workload}
This metric describes a mapping of workloads to the corresponding number of instances that are at least required to process that workload. Results obtained using this metric can be used to construct a scalability graph. 

\paragraph{Scalability Measurement Method}
In order to determine the required number of instances per workload, we determine for each workload $W$ the lowest number of instances that is able to process that workload $W$. Thus, for each tested number of instances $I$, we have to assess whether $I$ instances are sufficient to process workload $W$. Auxiliary Metric~1 can be used to answer that question.

\paragraph{Auxiliary Metric 1}
This metric provides a binary result whether $I$ instances are sufficient to process a workload $W$. There are two possible options to determine this.

\paragraph{Measurement Method 1.1}
Some stream processing engines provide strong guarantees regarding the time processing results are published as well as the amount of such published outputs. In such cases, we can determine a function which returns the amount of expected output messages for a set of input messages. 
Applying this function, we compare the amount of expected outputs with the amount of actually observed outputs. If both numbers match (possibly allowing for a certain tolerance), we conclude that the tested number of instances $I$ is sufficient to process workload $W$.
However, some stream processing engines such as Kafka Streams forward intermediate results of stateful operators. The amount of published intermediate results depends on the runtime behavior and, thus, the number of output results cannot be determined beforehand.

\paragraph{Measurement Method 1.2}
An alternative, indirect measurement method is to determine whether messages are queuing up before or while processing them. A tested amount of instances can be considered as sufficient if all generated messages are processed and no records are queuing up. Auxiliary Metric 2 determines whether records are persistently queuing up.

\paragraph{Auxiliary Metric 2}
This metric provides a binary result whether records are persistently queuing up between the workload generation and the SUT or within the SUT. Whereas temporary increases or decreases in the number of messages in queues are common (e.g, due to batching, shared resources, or varying network latency and throughput), we consider records to persistently queue up if the number of messages in queues increases over a larger period of time.
Depending on the capabilities of the SUT, this metric can either be measured directly or indirectly.

\paragraph{Measurement Method 2.1}

\begin{figure*}[tb]
	\begin{subfigure}[b]{0.49\linewidth}
		\centering
		\includegraphics[width=\textwidth]{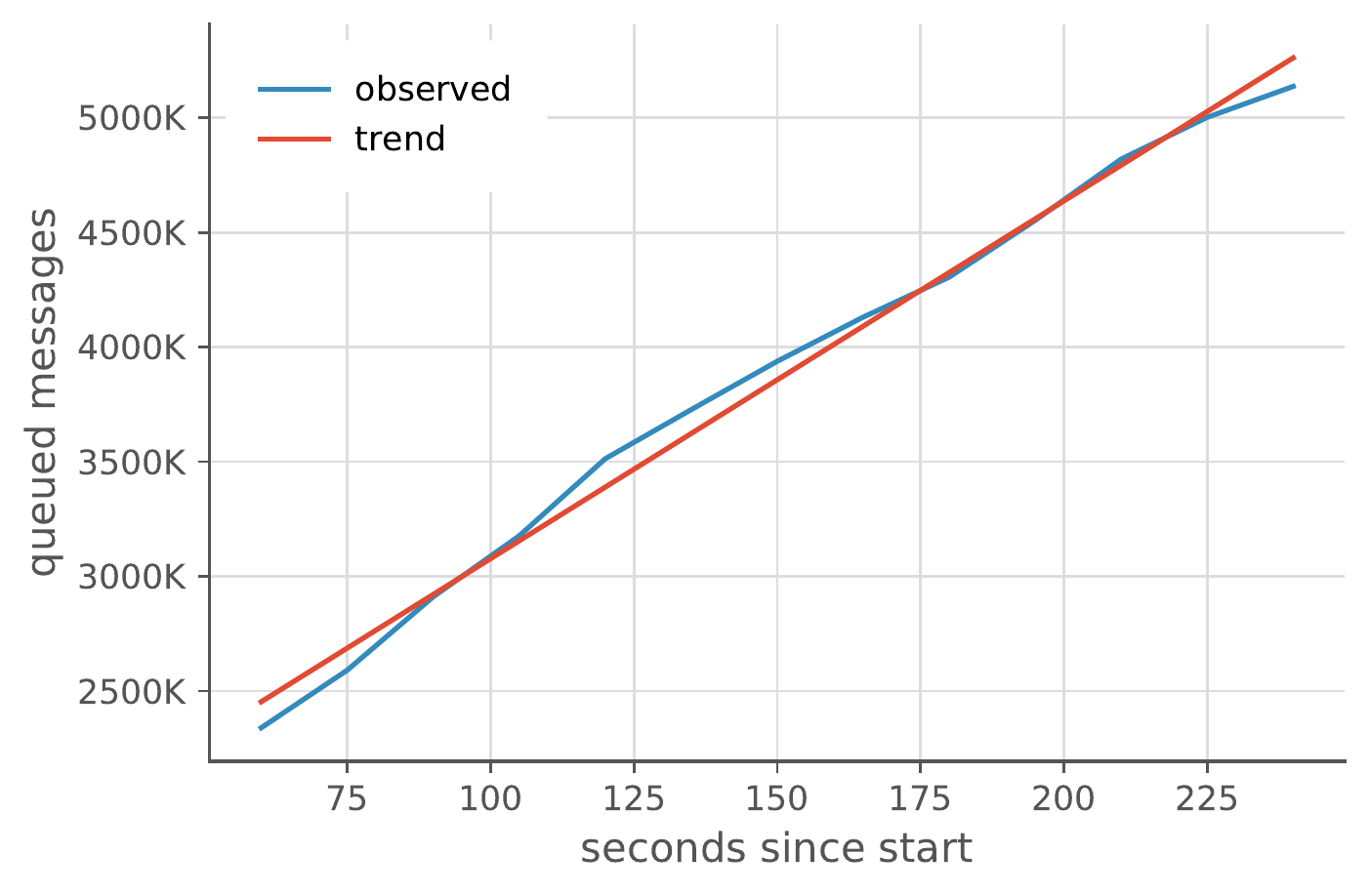}%
		\caption{Trend slope: 15605 msg./s}%
		\label{fig:lag-trend-increase}
	\end{subfigure}
	\hfill
	\begin{subfigure}[b]{0.49\linewidth}
		\centering
		\includegraphics[width=\textwidth]{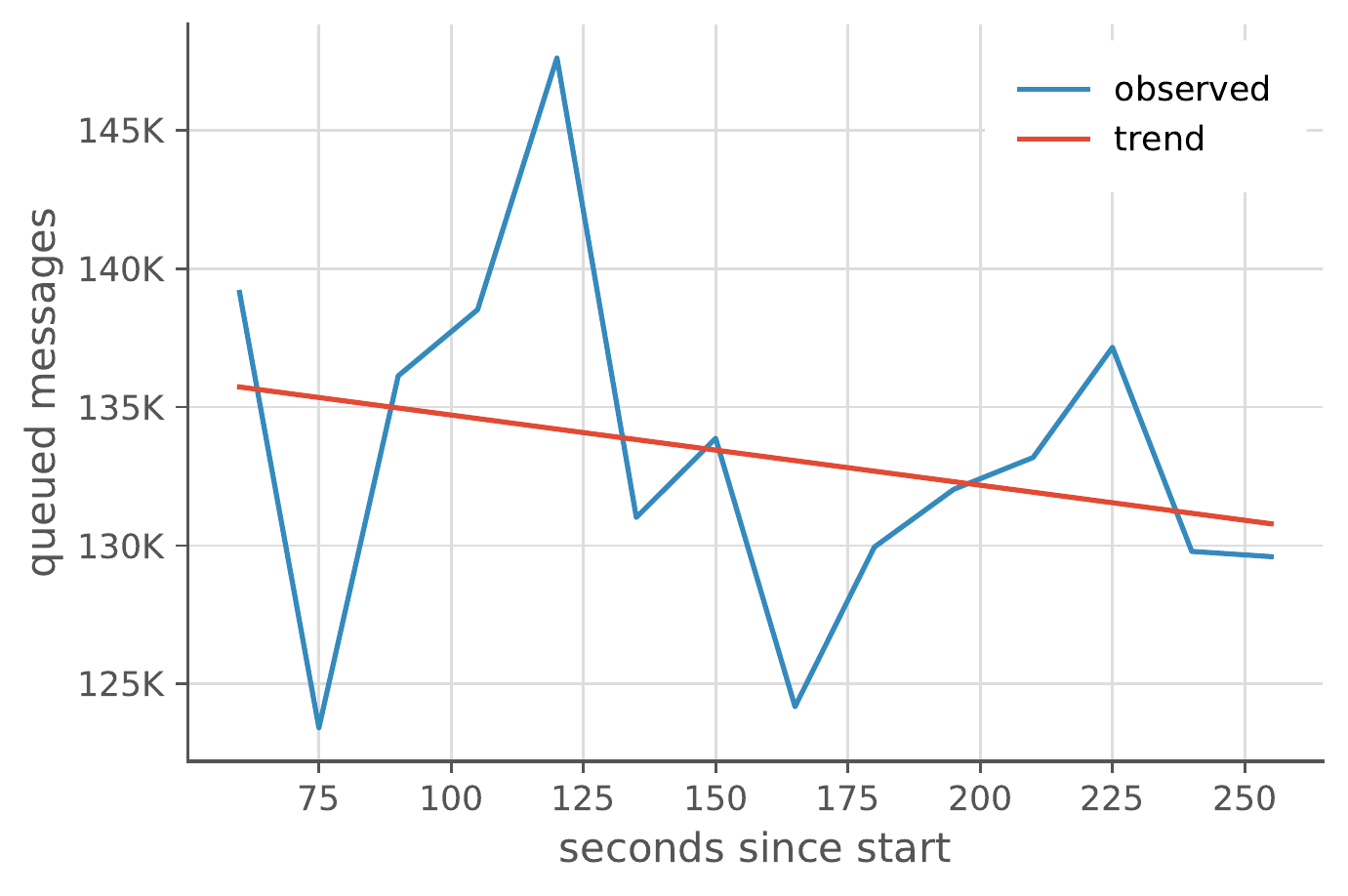}%
		\caption{Trend slope: $-25$ msg./s}%
		\label{fig:lag-trend-constant}
	\end{subfigure}
	\caption{Examples of recorded queue sizes over time and the trend line computed using linear regression.}
	\label{fig:lag-trend}
\end{figure*}

If the SUT provides an appropriate interface to monitor the number of records within queues, we track the number of queued messages over time and apply linear regression to fit a trend line (see Figure~\ref{fig:lag-trend}). The slope of this line indicates the average amount of messages per time unit by which the queues increase or decrease. For example, a slope of 1000 messages per second means that the number of messages in queues increases in average by 1000 messages per second.

Ideally, for a SUT providing sufficient resources, the trend's slope would be zero. However, since monitored queue sizes fluctuate considerably over time, linear regression usually calculates a slightly rising or falling trend line for these cases. Therefore, we consider records as persistently queuing up if the trend's slope is greater than a defined threshold. 
For example, with a threshold of 2000 messages per second, the records in Figure~\ref{fig:lag-trend-increase} are considered as persistently queuing up, whereas the number of instances processing records in Figure~\ref{fig:lag-trend-constant} is sufficient.

\paragraph{Measurement Method 2.2}
If a SUT does not provide information regarding the amount of records stored in queues, an alternative method to measure whether records are queuing up is to monitor the event-time latency \cite{Karimov2018} of messages. Event-time latency is the time passed between the time a message was created (event time) and the time the message's processing has been completed.

An increasing event-time latency over time means that messages are created faster than they can be processed. Thus, messages are queuing up either between workload generation and the SUT or within the SUT. Analogously to Method 2.1, we apply linear regression to fit a trend line of the event-time latency and consider records as persistently queuing up if the trend's slope is greater than a defined threshold.

\section{Benchmarking Framework Architecture}\label{sec:framework}

In this section, we present our framework architecture for executing scalability benchmarks. Our proposed benchmarking method in Section~\ref{sec:method} requires an individual benchmark per use case and workload dimension. Hence, our benchmarking framework can be configured by the following parameters:

\begin{enumerate}
	\item An implementation of the use case that should be benchmarked
	\item Configurations for the SUT including messaging system and execution environment
	\item The workload dimension, scalability should be benchmarked for
	\item A workload generator that generates workloads along the configured dimensions
	\item A list of workloads for the configured dimension to be tested
	\item A list of numbers of instances to be tested
\end{enumerate}

\begin{figure*}[t]
	\centering%
	\includegraphics[width=\textwidth]{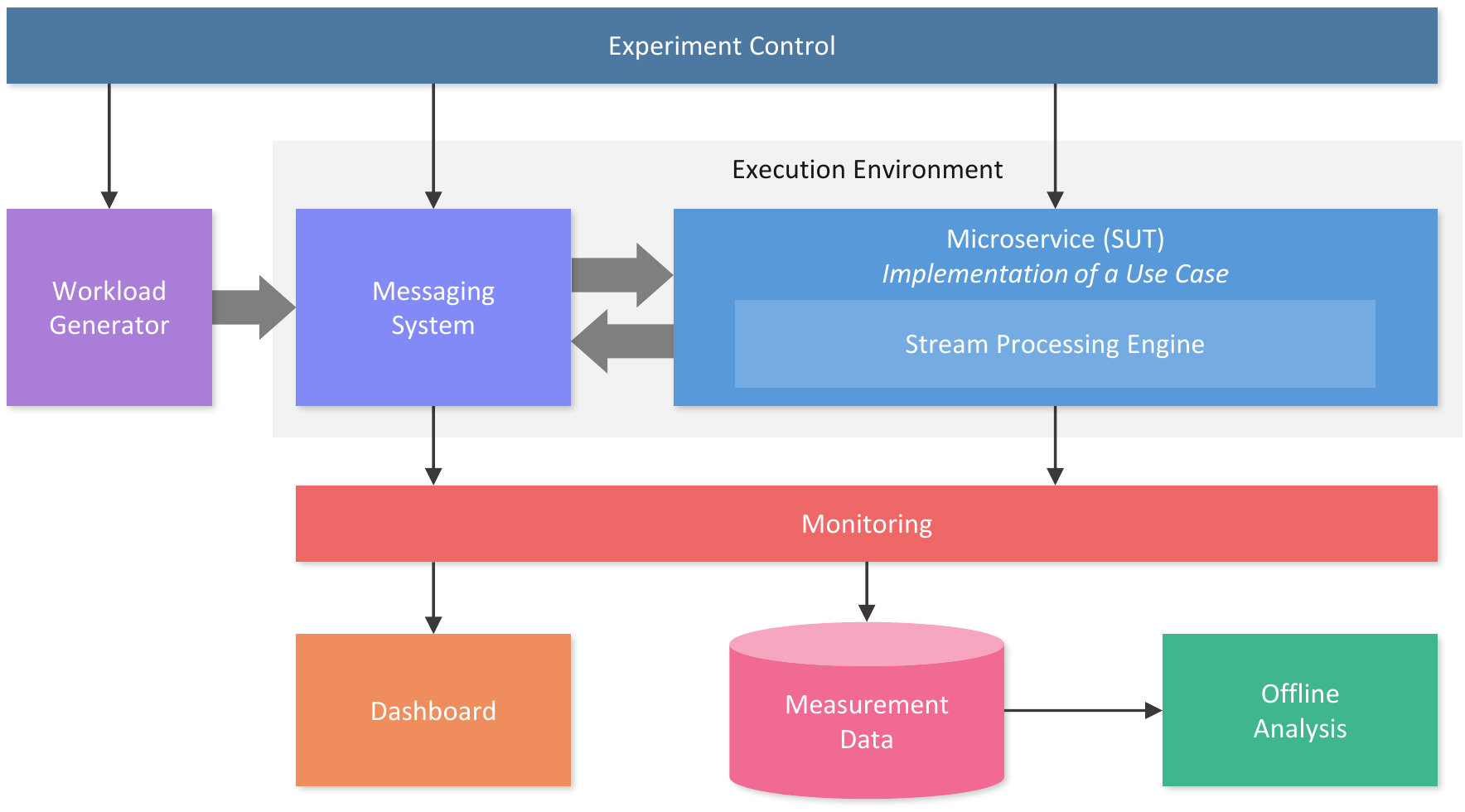}%
	\caption{The Theodolite framework architecture for executing scalability benchmarks.}
	\label{fig:execution-architecture}
\end{figure*}%

Following our defined metrics (Section~\ref{sec:metrics}), our benchmarking framework conducts subexperiments for each tested workload with each tested number of instances. For each subexperiment, it determines whether the currently tested number of instances is sufficient to process the currently tested workload.
Figure~\ref{fig:execution-architecture} depicts our architecture for executing scalability benchmarks. It consists of the following components:

\paragraph{Experiment Control}
The central experiment control is started at the beginning of each scalability benchmark and runs throughout its entire execution. 
For each subexperiment, it starts and configures the workload generator component to generate the current workload of the tested dimension. Further, it starts and replicates the SUT according to the evaluated number of instances.
After each subexperiment, this component resets the messaging system, ensuring no queued data can be accessed by the following subexperiment. 

\paragraph{Workload Generator}
This component generates a configurable constant workload of a configurable workload dimension. It fulfills the function of a data source in a big data streaming system, such as an IoT device or another microservice.
Since different use cases require different data input formats, we envisage individual workload generators per use case. However, individual workload generators can share large parts of their implementations.

\paragraph{Messaging System}
In event-driven, microservice-based architectures, individual services usually communicate with each other via a dedicated messaging system. Our benchmarking architecture therefore contains such a system, serving as a message queue between workload generator and stream processing engine and as a sink for processed data.
State-of-the-art messaging systems already partition the data for the stream processing engine and are, thus, likely to have high impact on the engine's scalability. They provide plenty of configuration options, making it reasonable to benchmark different configurations against each other.

\paragraph{Microservice (SUT)}
This component acts as a microservice that applies stream processing and, thus, is the actual SUT. This microservice fulfills a specific use case, such as those presented in Section~\ref{sec:UseCases}.
An implementation of this microservice uses a certain stream processing engine along with a certain configuration, which should be benchmarked.
The stream processing engine receives all data to be processed from the messaging system and, optionally, writes processing results back to it.

\paragraph{Monitoring}
The monitoring component collects runtime information from both the messaging system and the stream processing engine. This includes data to be displayed by the dashboard and data required to actually measure the scalability of the SUT.

\paragraph{Dashboard}
Our proposed architecture contains a dashboard for observation of benchmark executions. It visualizes monitored runtime information of the execution environment, the messaging system, and the SUT. Thus, it allows to verify the experimental setup (e.g., number of deployed instances and number of generated messages per seconds).

\paragraph{Offline Analysis}
Based on the raw monitoring data, a dedicated component evaluates the scalability of the SUT by computing the required metrics as described in Section~\ref{sec:metrics}. This component is executed offline after completing all subexperiments. Since we store monitoring data persistently, we can repeat all computations at any time without re-executing the underlying experiments.

\section{Cloud-native Implementation}\label{sec:implementation}

In this section, we present our open-source implementation\footnote{\url{https://github.com/cau-se/theodolite}}
of our proposed scalability benchmarking framework.
As our benchmarks represent real use cases, also their execution environment should correspond to that of real deployments.
Microservices are increasingly used as building blocks of cloud-native applications \cite{Hasselbring2018}.
Hence, our implementation deploys all benchmark components as containers in a cloud environment, orchestrated by Kubernetes \cite{Burns2016}. This includes the workload generator, the messaging system, and the stream processing engine as well as the monitoring and dashboard components. Thus, the only two requirements for arbitrary workload generators and use case implementations are that they are (1) deployed as containers and (2) support sending and receiving messages from the selected messaging system.

We select Apache Kafka \cite{Kreps2011} as the messaging system for our implementation. 
Kafka is widely used in industry as a messaging system connecting microservices.\footnote{\url{https://kafka.apache.org/powered-by}} It is a supported data source for most stream processing engines and, thus, heavily used in stream processing benchmarks \cite{Lu2014, Chintapalli2016, Lopez2016, Hesse2018, Karakaya2017, Nasiri2019, Pagliari2020, Shahverdi2019, Bordin2020, VanDongen2020}.
Between the execution of individual subexperiments, the experiment control resets Kafka for the next subexperiment. It recreates all required Kafka topics according to benchmark configurations such as partition count or replication factor.

\begin{figure*}[t]
	\centering%
	\includegraphics[width=\linewidth]{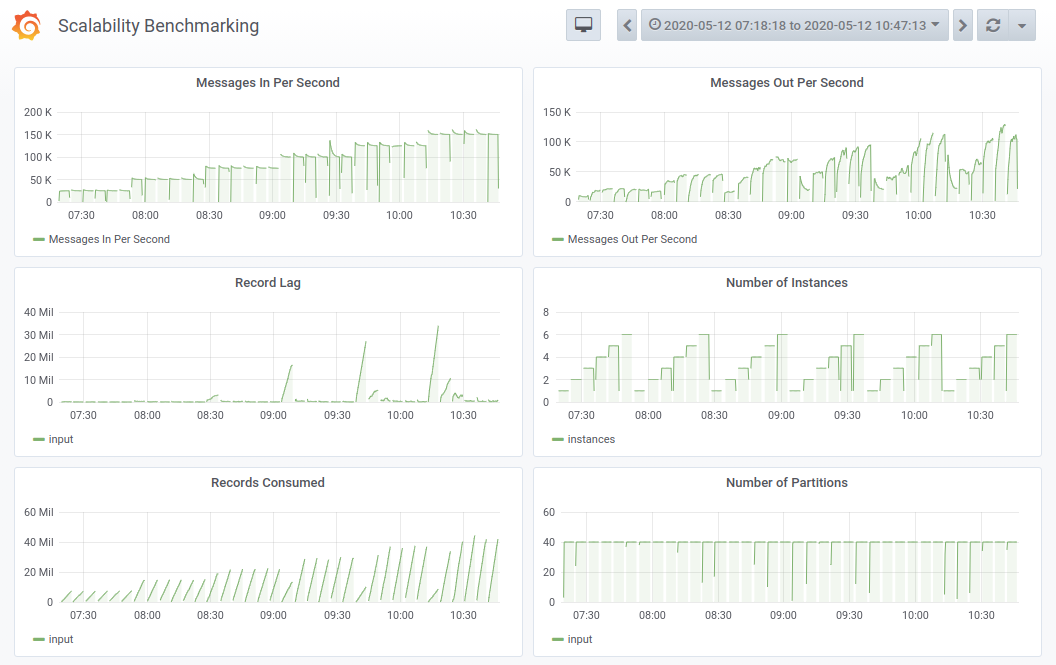}%
	\caption{Screenshot of our dashboard for observing benchmark executions.}
	\label{fig:screenshot-dashboard}
\end{figure*}

In our implementation we apply Measurement Method 1.2 and 2.1 of Section~\ref{sec:metrics}, which requires to continuously monitor the number of queued messages.
Kafka consumers provide a \textit{record lag} metric, stating the difference between ever-appended and already consumed messages of a topic.
The stream processing engine exposes this metric, optionally augmented with the amount of internally queued messages, to be recorded by the monitoring component.
Kafka provides additional metrics such as numbers of messages written to topics per second.
We use Prometheus\footnote{\url{https://prometheus.io}} to collect these metrics as well as performance and state metrics from Kubernetes and visualize them in a Grafana\footnote{\url{https://grafana.com/grafana/}} dashboard (see Figure~\ref{fig:screenshot-dashboard}). 
Prometheus and Grafana are widely used tools for monitoring cloud-native applications \cite{Casalicchio2017}.

After each execution of a subexperiment, the monitored consumer lag time series is stored in a comma-separated values (CSV) file, allowing for later offline analysis and sharing experimental raw results.
For offline analysis of an executed benchmark, we provide a Jupyter\footnote{\url{https://jupyter.org}} notebook that loads all time series of subexperiments, computes the number of required instances per workload, and creates a scalability graph for the benchmarked SUT.
This notebook can easily be adjusted, supporting evaluations of different parameters.

Along with our benchmarking framework, we provide four benchmarks, one for each use case of Section~\ref{sec:UseCases}. For each use case, we provide implementations with Apache Flink \cite{Carbone2015}, a stream processing engine widely used in research and production (see Section~\ref{sec:Related}), and Kafka Streams \cite{Sax2018}, a stream processing framework built upon Apache Kafka's capabilities for reprocessing distributed, replicated logs \cite{Wang2015}.
For each implemented benchmark, we also provide a corresponding workload generator.

\section{Experimental Evaluation}\label{sec:Evaluation}

In this section, we apply our proposed scalability benchmarking method to evaluate the effect of different deployment options on scalability. For this purpose, we use our proposed benchmarking framework along with our presented implementation and select Kafka Streams and Flink as stream processing engines. For Kafka Streams, we evaluate scalability regarding the number of topic partitions in Kafka, Kafka Streams' commit interval, and the resources provided from Kubernetes for individual Kafka Streams instances. For Flink, we evaluate scalability regarding different options for checkpointing. Furthermore, we compare the scalability results of Kafka Streams and Flink. For each deployment option, we benchmark the four use cases described in Section~\ref{sec:UseCases}.

For the use cases UC1, UC2, and UC3, we benchmark scalability along the workload dimension of increasing amounts of different keys. For use case UC4, we benchmark scalability along the dimension of numbers of nested groups in the aggregation hierarchy. We generate data only for those sensors, which are specified in the hierarchy. Hence, scaling with the number of nested groups effectively corresponds also to scaling with the number of different keys.
Partitioning data streams based on keys is the fundamental concept of most stream processing engines for scalability (see Section~\ref{sec:stream-processing-foundations}). Thus, evaluating scalability with the number of different keys is a natural choice for executing the first scalability benchmarks.

\subsection{Experiment Setup}

All experiments are executed in a private cloud, operated by Kubernetes 1.14. It provides 4 nodes, each equipped with 384\,GB RAM and $2\times16$ CPU cores, providing 128~cores in total (hyper-threading disabled). All nodes are connected via 10\,Gbit/s Ethernet.
We deploy 10 Kafka brokers and, unless otherwise stated, use the following configurations (see Table~\ref{tab:evaluation-options}).
Each topic involved in our benchmark is configured with 40~partitions.
Kafka, Kafka Streams, and Flink are deployed with their default configuration, except that we set Kafka Stream's commit interval to 100\,ms to reduce the latency of outputs.
In particular, this means that each Kafka Streams instance and each Flink task manager only spawns one processing thread.
Each processing instance (Kafka Streams application instance or Flink task manager) is deployed as an individual pod, restricted to use 1000\,\textit{milliCPUs} (i.e., one CPU core) and 4\,GB memory.
For our benchmarks of Flink, we additionally deploy one job manager as a pod.

\begin{table}
	\centering
	\caption{Overview of evaluated deployment options.}
	\label{tab:evaluation-options}
	\small
	\begin{tabular}{lcc}
		\toprule 
		Deployment Option & Default Value & Evaluated Values \\
		\midrule 
		Kafka Partition Count & 40 & 40, 160, 400, 1600 \\
		Kafka Streams Commit Interval & 100\,ms & 10\,ms, 100\,ms \\
		Flink Checkpointing & disabled & disabled, 100\,ms, 10\,s \\
		CPU cores per Kubernetes pod & 1.0 & 0.5, 1.0, 2.0 \\
		Memory per Kubernetes pod & 4\,GB & 2\,GB, 4\,GB, 8\,GB \\
		\bottomrule
	\end{tabular}
\end{table}

We deploy up to 100~processing instances to avoid our available hardware become the bottleneck. Moreover, we closely observe Kubernetes monitoring metrics throughout all experiments to ensure that we do not hit any limits regarding CPU, memory, network or disk usage of the individual nodes.
The tested workloads are generated by up to 4~instances, depending on the load that should be generated.

Following our proposed benchmarking framework from Section~\ref{sec:framework}, we execute one subexperiment for each configured number of processing instances with each configured workload. Each subexperiment is executed for 5~minutes, where we consider the first minute as warm-up period and only analyze how the record lag evolves after the first minute. We consider messages as persistently queuing up if the computed trend line of the record lag has a slope that exceeds 2000~messages per second.
In preparatory experiments, we determined that these settings lead to stable results. However, to increase the general validity of our results further evaluations are required (see Section~\ref{sec:threats-to-validity}).

In all experiments, we generate one message per second and simulated data source.
For the individual benchmarks we apply the following configurations:
As suggested in Section~\ref{sec:UseCases}, use case UC1 does not include storing records to a real database as such a database would likely become the bottleneck for these benchmarks.
Use case UC2 is configured with an aggregation time window size of one minute.
In use case UC3, the time window size is set to 3 days with starting a new window every 24 hours, resulting in 3 overlapping windows. The time attribute, for which data is aggregated, is the hour of day, which ultimately means that this configuration computes summary statistics for each hour of day over the last 3 days.
Use case UC4 is configured to use tumbling window aggregations. The aggregation hierarchy is configured with 4~elements per group, resulting in $4^n$~total sensors for $n$~nested groups.

\subsection{Results and Discussion}

In the following, we present and discuss the results of our individual evaluations.
Our replication package and the collected data of our experiments are published as as supplemental material \cite{ReplicationPackage}, such that other researchers may repeat and extend our work. Note that in all experiments, the amount of data sources corresponds to the throughput in messages per second as we generate one message per data source and second.

\subsubsection{Kafka Partition Count}

We evaluate the effect of different numbers of partitions of Kafka topics on Kafka Streams' scalability. The amount of partitions controls the maximal parallelism of Kafka Streams applications. However, too many partitions are likely to cause significant overhead, suggesting to set the number of partitions as high as required, but as low as possible.
We evaluate how scalability behaves with 40, 160, 400, and 1600~partitions for each input topic. Note that Kafka Streams uses the same number of instances for internal repartitioning topics.

\begin{figure*}[tb]
	\begin{subfigure}[b]{0.49\textwidth}
		\centering
		\includegraphics[width=\textwidth]{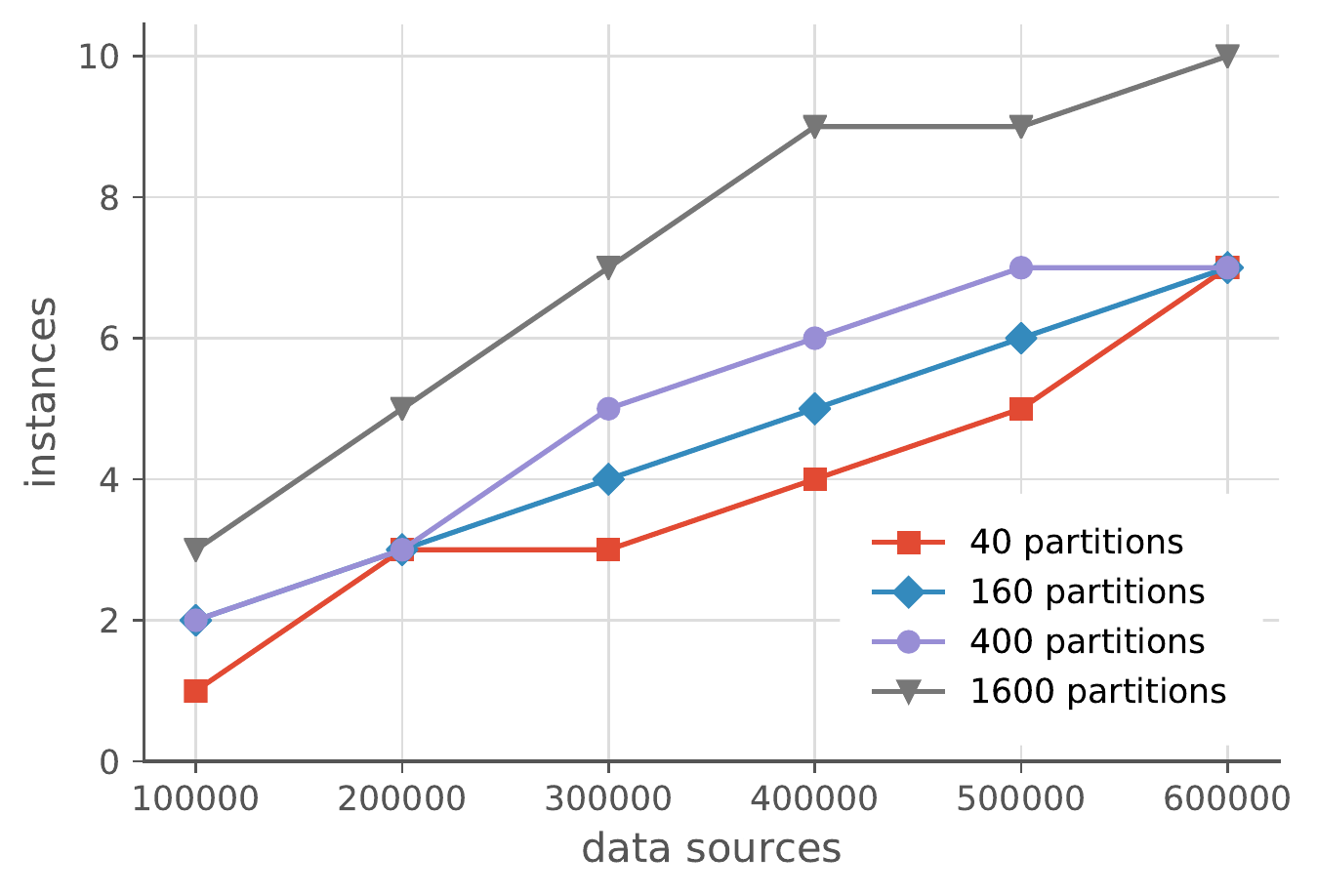}%
		\caption{Use case UC1: Database Storage}
	\end{subfigure}
	\hfill
	\begin{subfigure}[b]{0.49\textwidth}
		\centering
		\includegraphics[width=\textwidth]{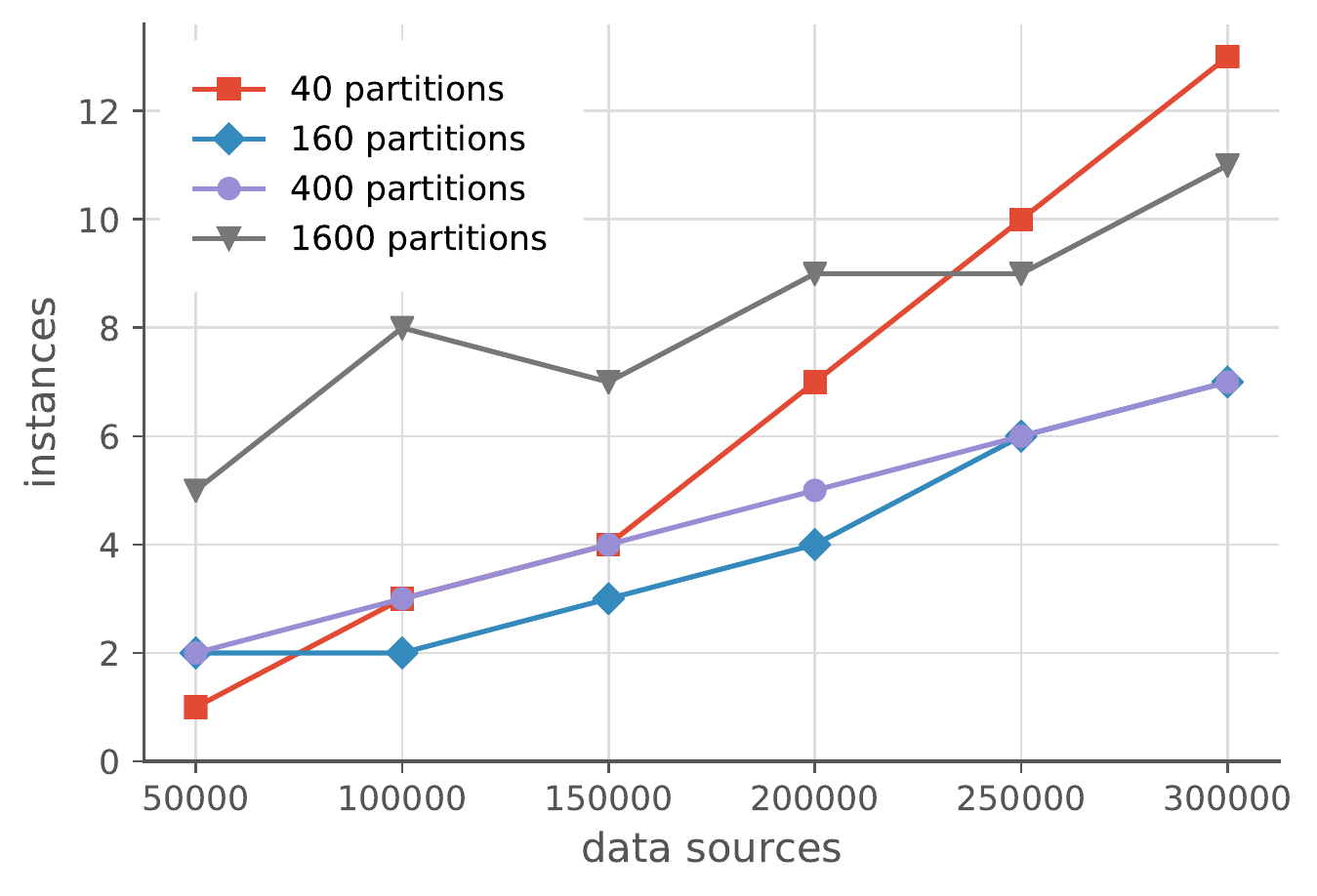}%
		\caption{Use case UC2: Downsampling}
	\end{subfigure}
	
	\vspace{1em}
	
	\begin{subfigure}[b]{0.49\textwidth}
		\centering
		\includegraphics[width=\textwidth]{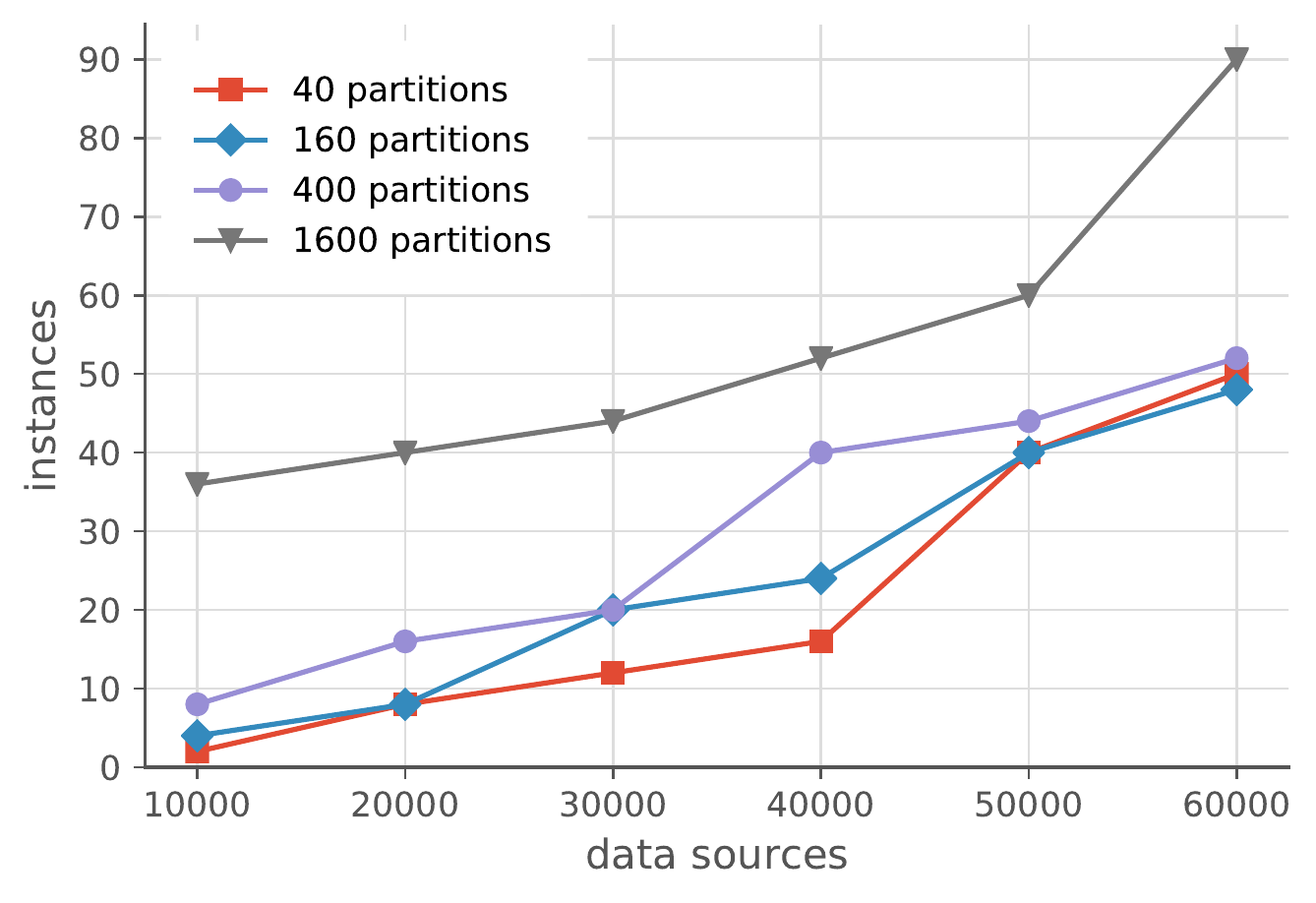}%
		\caption{Use case UC3: Aggregating Time Attributes}
	\end{subfigure}
	\hfill
	\begin{subfigure}[b]{0.49\textwidth}
		\centering
		\includegraphics[width=\textwidth]{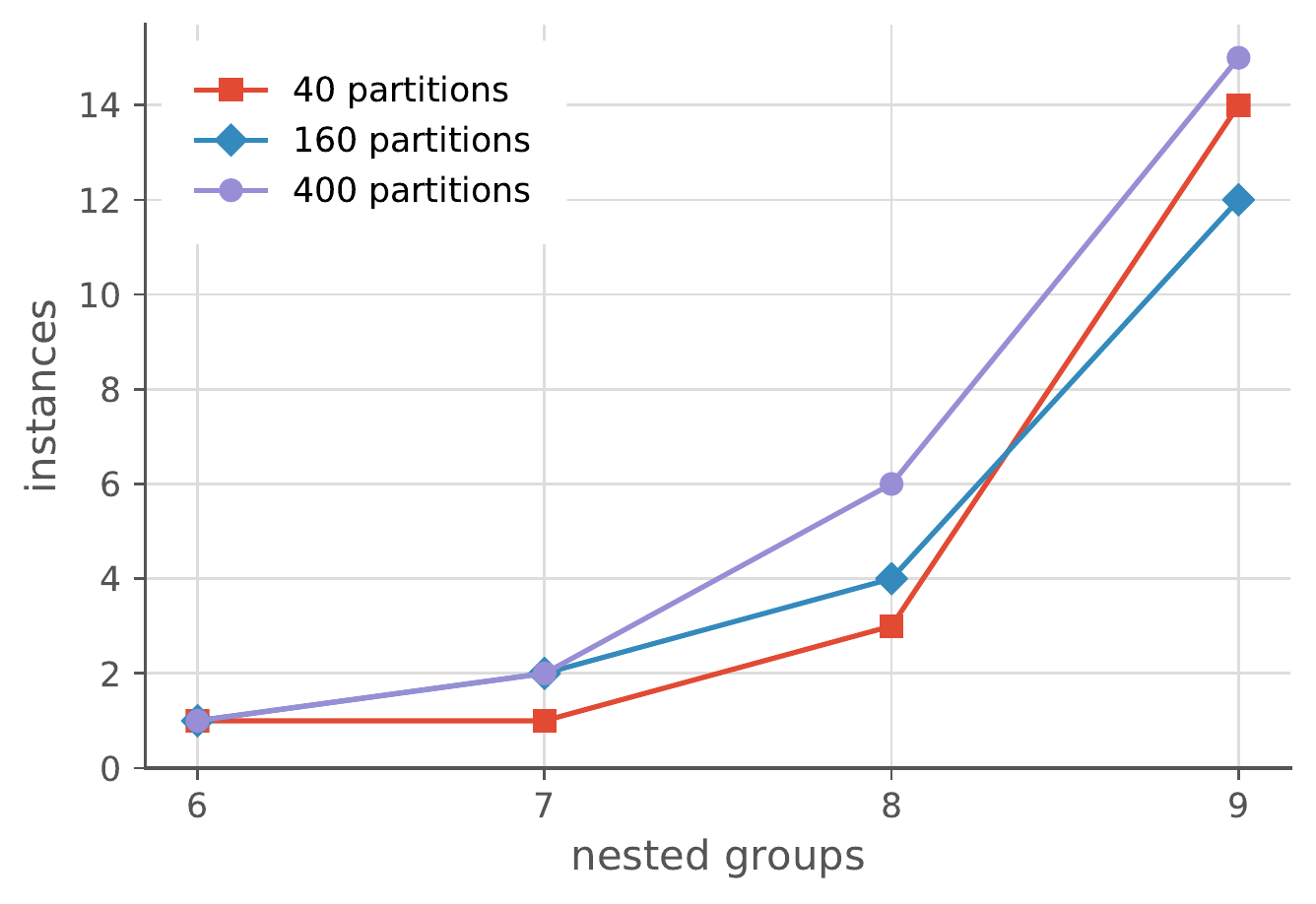}%
		\caption{Use case UC4: Hierarchical Aggregation}
	\end{subfigure}
	\caption{Kafka Streams scalability benchmark results for different partition counts of Kafka topics.}
	\label{fig:eval-partitions}
\end{figure*}

Figure~\ref{fig:eval-partitions} shows the results of our experiments for the different partition counts we evaluate. Apart from some fluctuations, we observe that the amount of required Kafka Streams instances increases linearly for use cases UC1 and UC2 with increasing amounts of data sources. The amount of required instances for use case UC4 increases in an exponential fashion.
As also the amount of data sources increases exponentially with increasing the number of nested groups, our results suggest that Kafka Streams scales also linearly for use case UC4 with the number of data sources.
Our implementation of use case UC4 does not process any data with 1600 partitions.

We observe that the number of required instances for use case UC3 is significantly higher than for similar workloads on the other use cases.
Moreover, in contrast to the other use cases, the amount of required instances rises steeper than linearly.
This could be either due to the characteristics of UC3 or the large number of instances required for processing.
Use case UC3's deployment with 40 partitions is able to use more than 40 instances. This is due to the fact that Kafka Streams performs a repartitioning and, thus, creates two streaming operators per partition.
	
In general, we observe that using more partitions requires more instances for the same workload.
However, the more processing instances are deployed the smaller are the advantages of having less partitions. In the case of use case UC2, fewer partitions for large workloads perform even worse.
Our experiments show that Kafka Streams scales independent of the chosen partition count.
Nevertheless, we conclude that it is still an important configuration parameter for a resource-efficient deployment.
This is also emphasized by the fact that the number of partitions can hardly be altered at runtime.

\subsubsection{Kafka Streams Commit Interval}

Kafka Streams' commit interval configuration specifies how often the current processing position at data streams is committed.
Effectively, it also controls how often intermediate results of stateful operations such as aggregations are forwarded. It, thus, has a huge impact on the event-time latency of stream processing application. We benchmark how scalability is affected by setting the commit interval to 10\,ms and 100\,ms.

\begin{figure*}[tb]
	\begin{subfigure}[b]{0.49\textwidth}
		\centering
		\includegraphics[width=\textwidth]{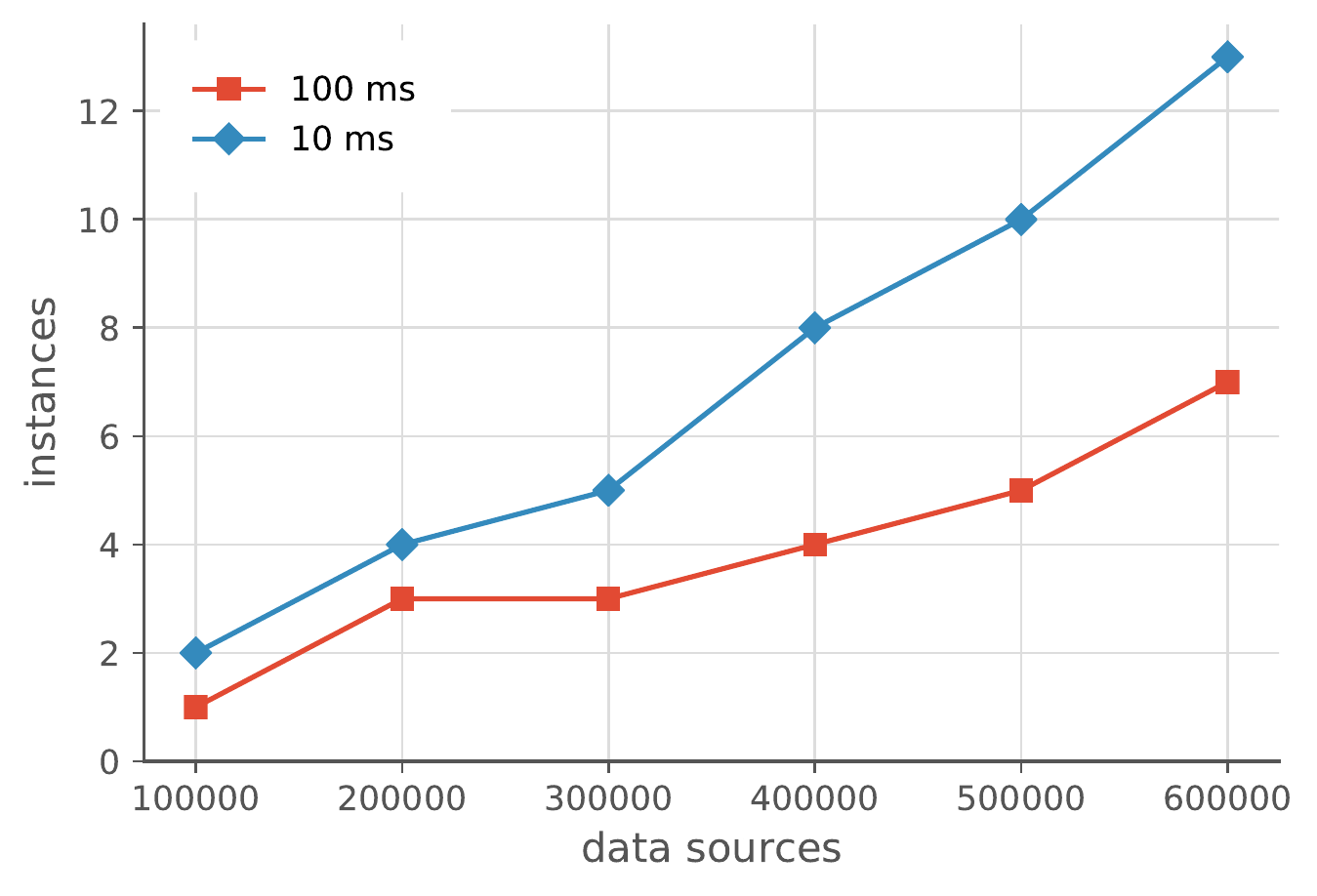}%
		\caption{Use case UC1: Database Storage}
	\end{subfigure}
	\hfill
	\begin{subfigure}[b]{0.49\textwidth}
		\centering
		\includegraphics[width=\textwidth]{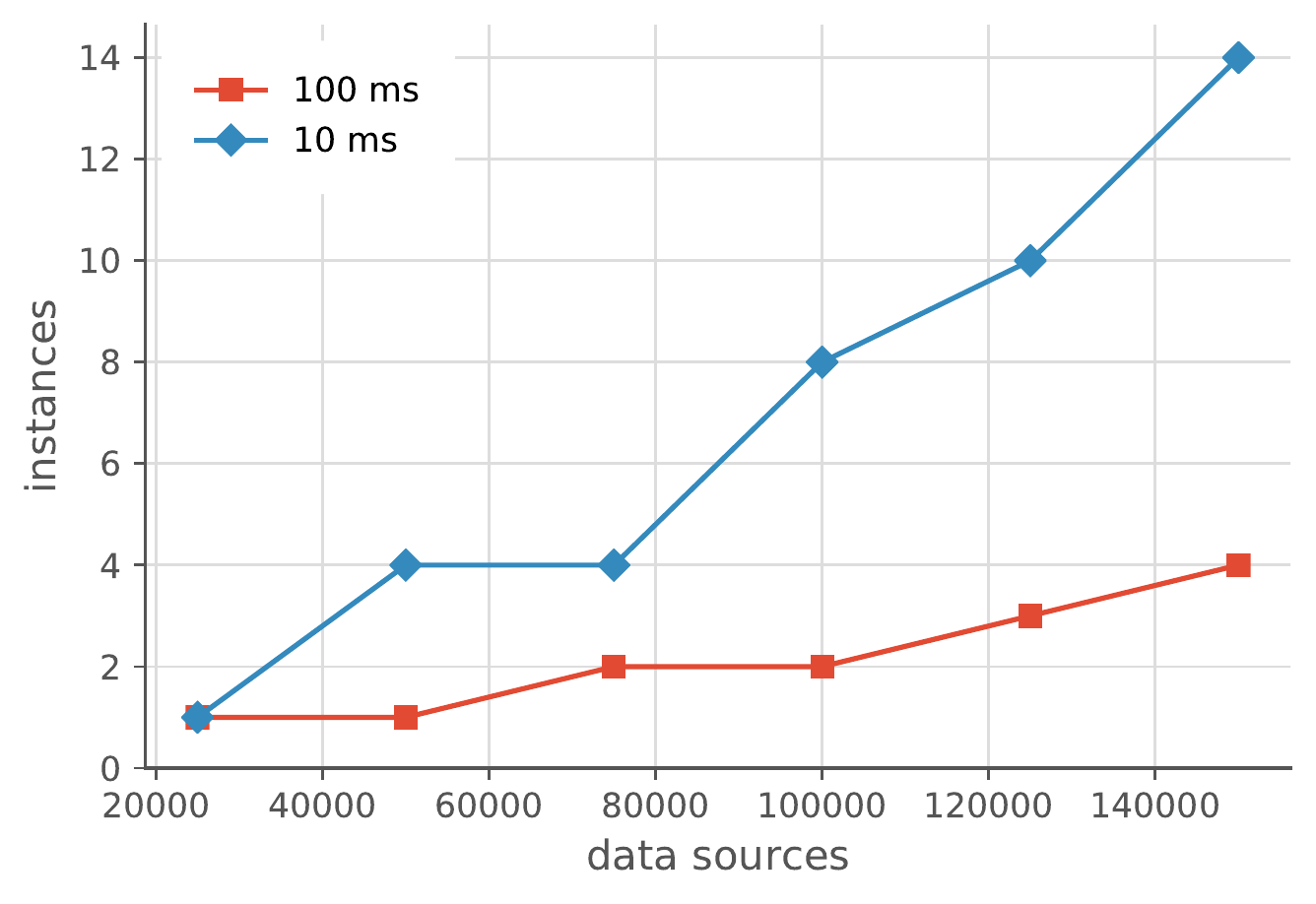}%
		\caption{Use case UC2: Downsampling}
	\end{subfigure}
	
	\vspace{1em}
	
	\begin{subfigure}[b]{0.49\textwidth}
		\centering
		\includegraphics[width=\textwidth]{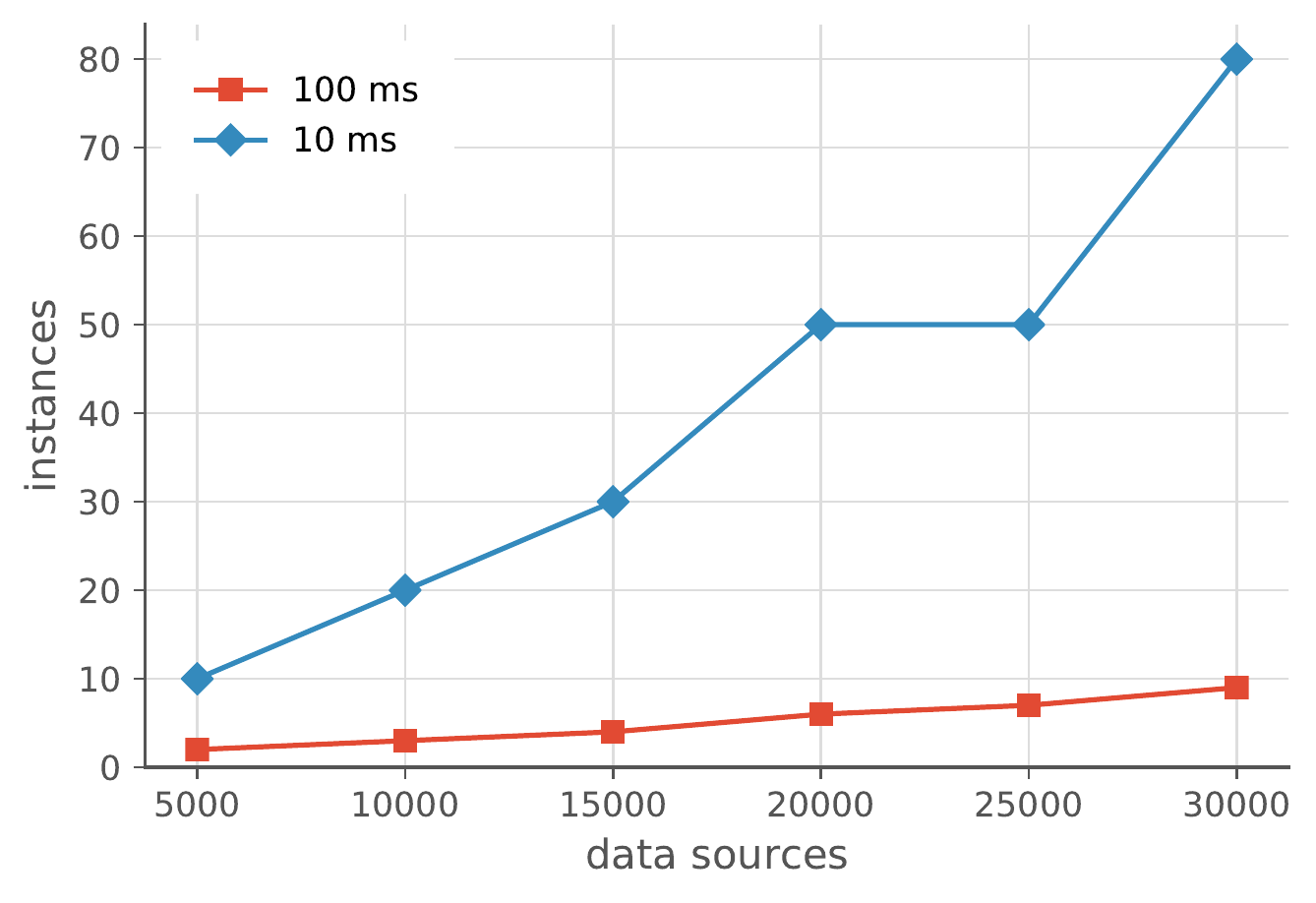}%
		\caption{Use case UC3: Aggregating Time Attributes}
	\end{subfigure}
	\hfill
	\begin{subfigure}[b]{0.49\textwidth}
		\centering
		\includegraphics[width=\textwidth]{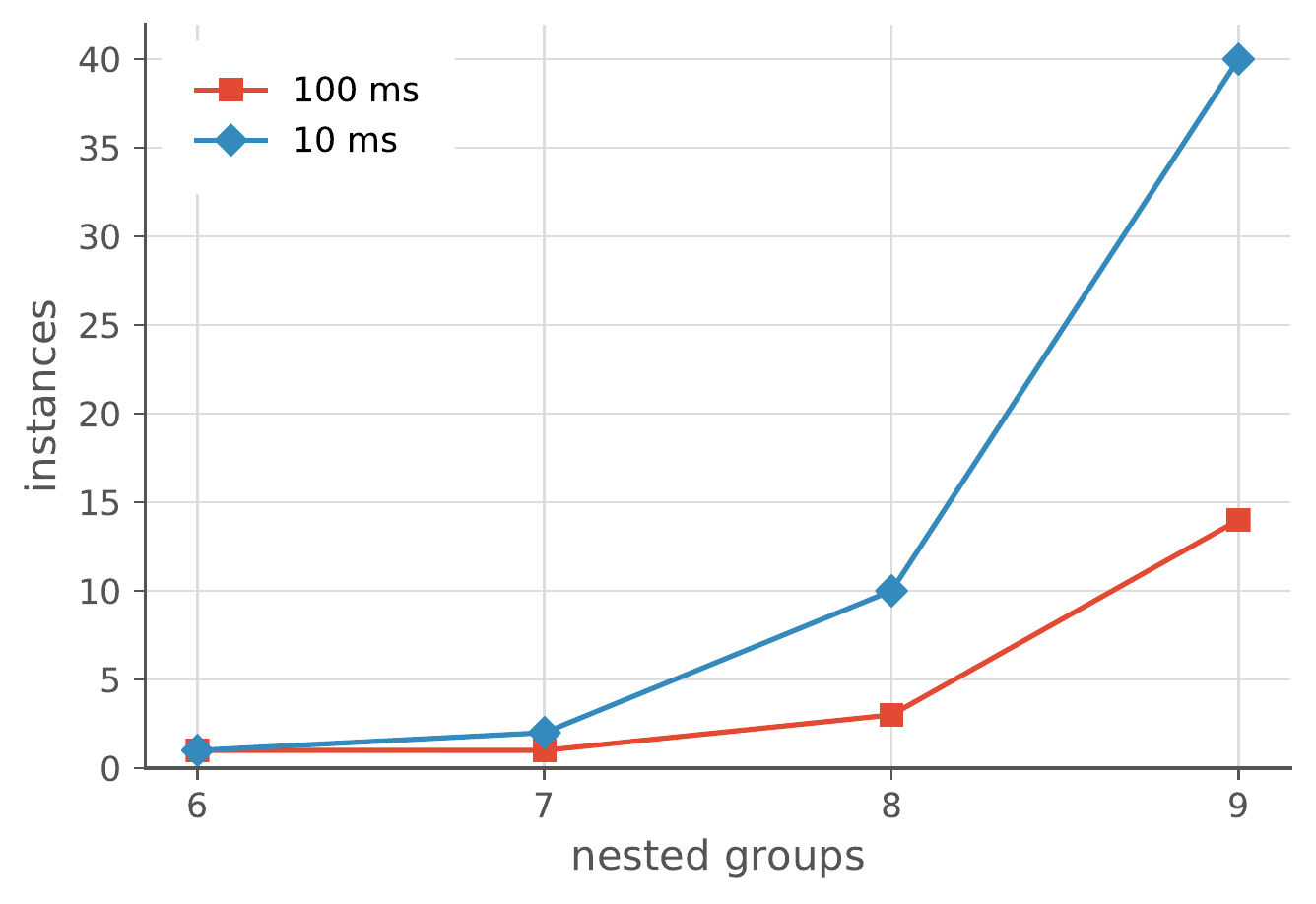}%
		\caption{Use case UC4: Hierarchical Aggregation}
	\end{subfigure}
	\caption{Kafka Streams scalability benchmark results for different Kafka Streams commit intervals.}
	\label{fig:eval-commit-intv}
\end{figure*}

Figure~\ref{fig:eval-commit-intv} shows the results of our experiments with different Kafka Streams commit intervals. 
The experiments for all use cases reveal that significantly more instances are required when using a lower commit interval.
The commit interval can thus be considered as a crucial configuration option, which should be set carefully.
It is remarkable that even for use case UC1, which is stateless, a lower commit interval causes significantly higher resource demands.
Nevertheless, our experiments lead us to conclude that Kafka Streams scales linearly for both evaluated commit intervals. Thus if very low latencies are required, Kafka Streams can be configured with a short commit interval at the costs of requiring significantly more computing resources.

\subsubsection{Provided Kubernetes Resources}

Kubernetes allows to restrict the resource usage of pods such as CPU cores and memory. With these benchmarks, we compare how scalability behaves when doubling or halving the resource restriction of Kafka Streams instances. We evaluate restrictions to 0.5, 1, and 2~CPU cores as well as 2, 4, and 8\,GB memory. Note that even when limiting the CPU to more than one core, each instance is still configured to use only one processing thread.

\begin{figure*}[tb]
	\begin{subfigure}[b]{0.49\textwidth}
		\centering
		\includegraphics[width=\textwidth]{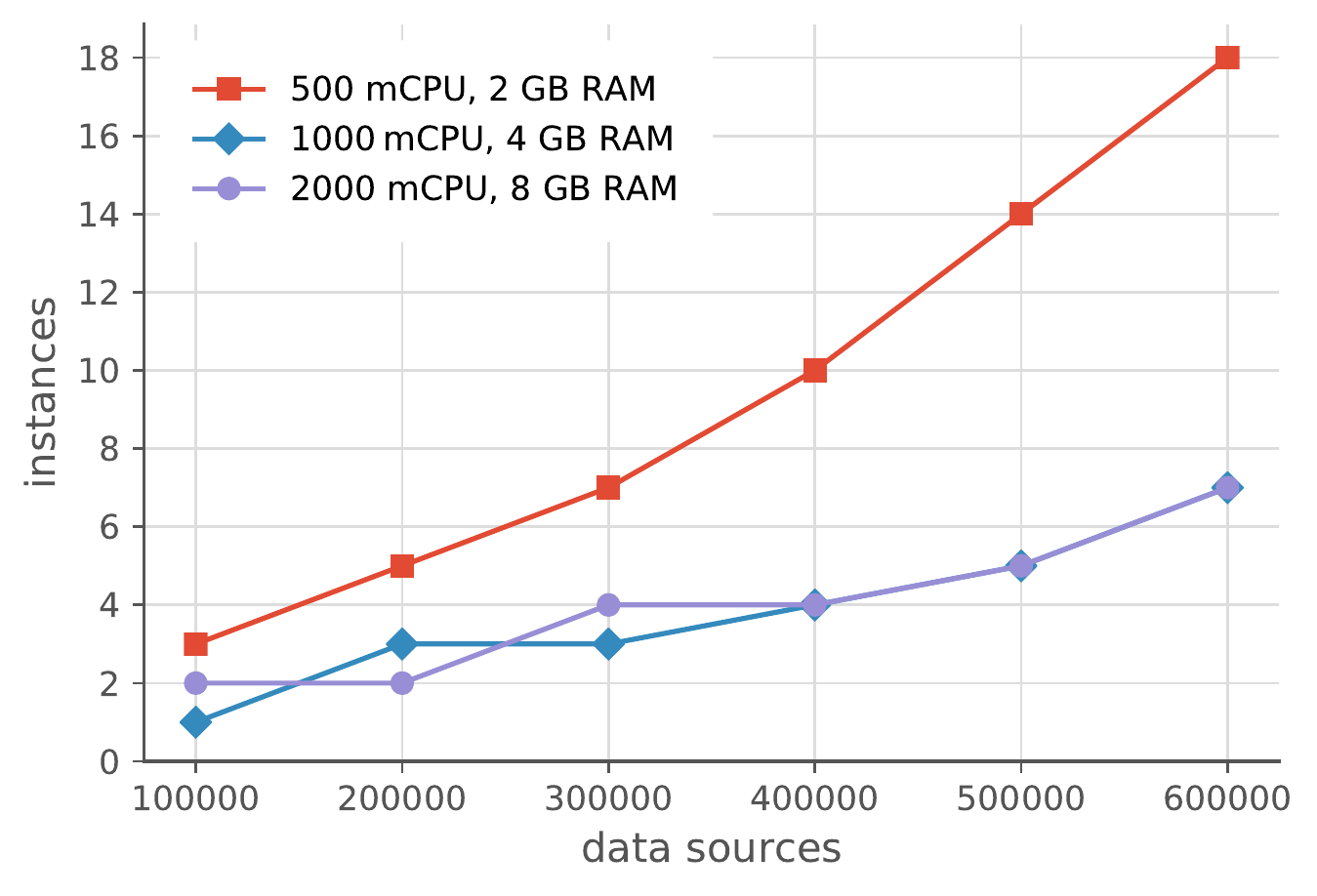}%
		\caption{Use case UC1: Database Storage}
	\end{subfigure}
	\hfill
	\begin{subfigure}[b]{0.49\textwidth}
		\centering
		\includegraphics[width=\textwidth]{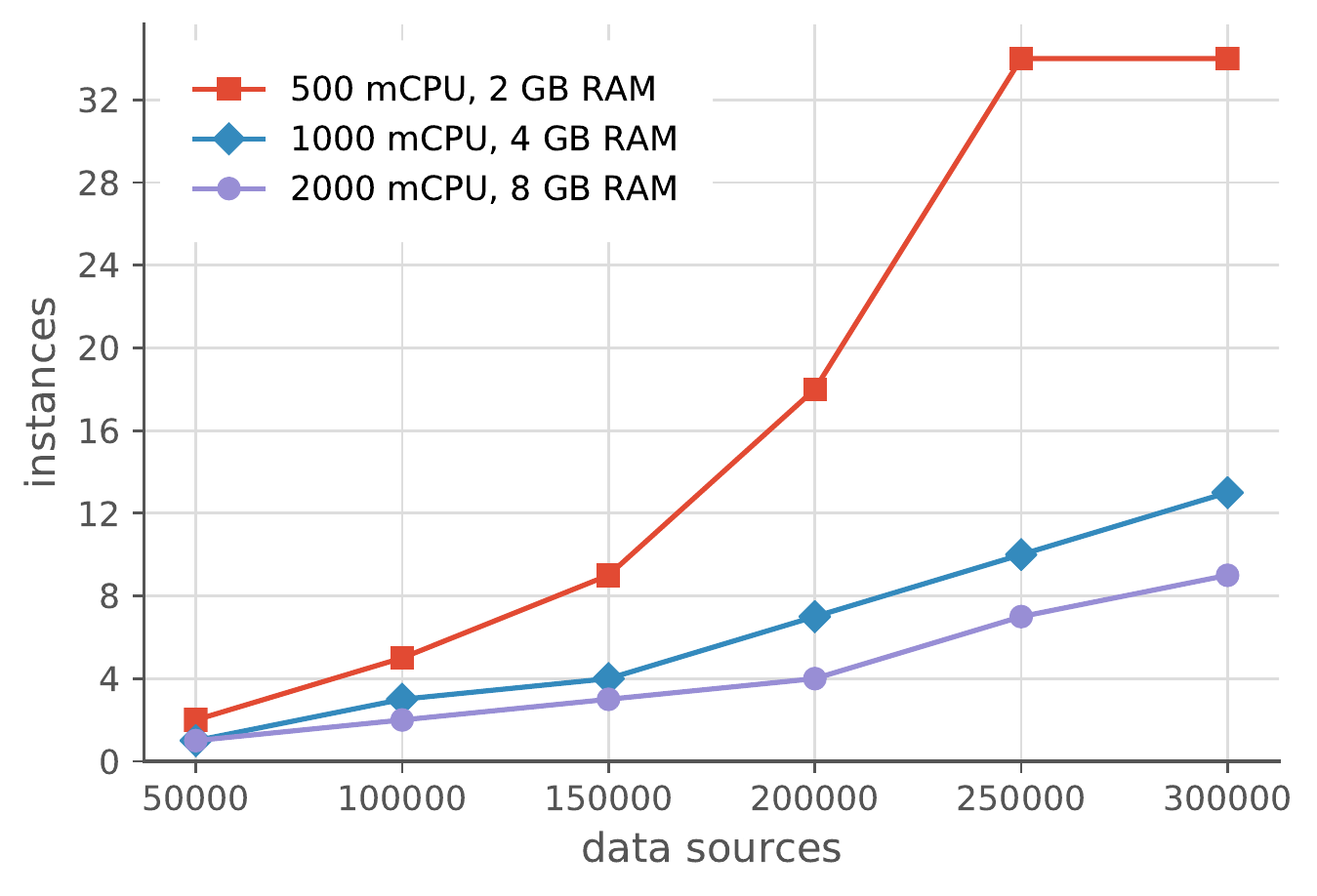}%
		\caption{Use case UC2: Downsampling}
	\end{subfigure}
	
	\vspace{1em}

	\begin{subfigure}[b]{0.49\textwidth}
		\centering
		\includegraphics[width=\textwidth]{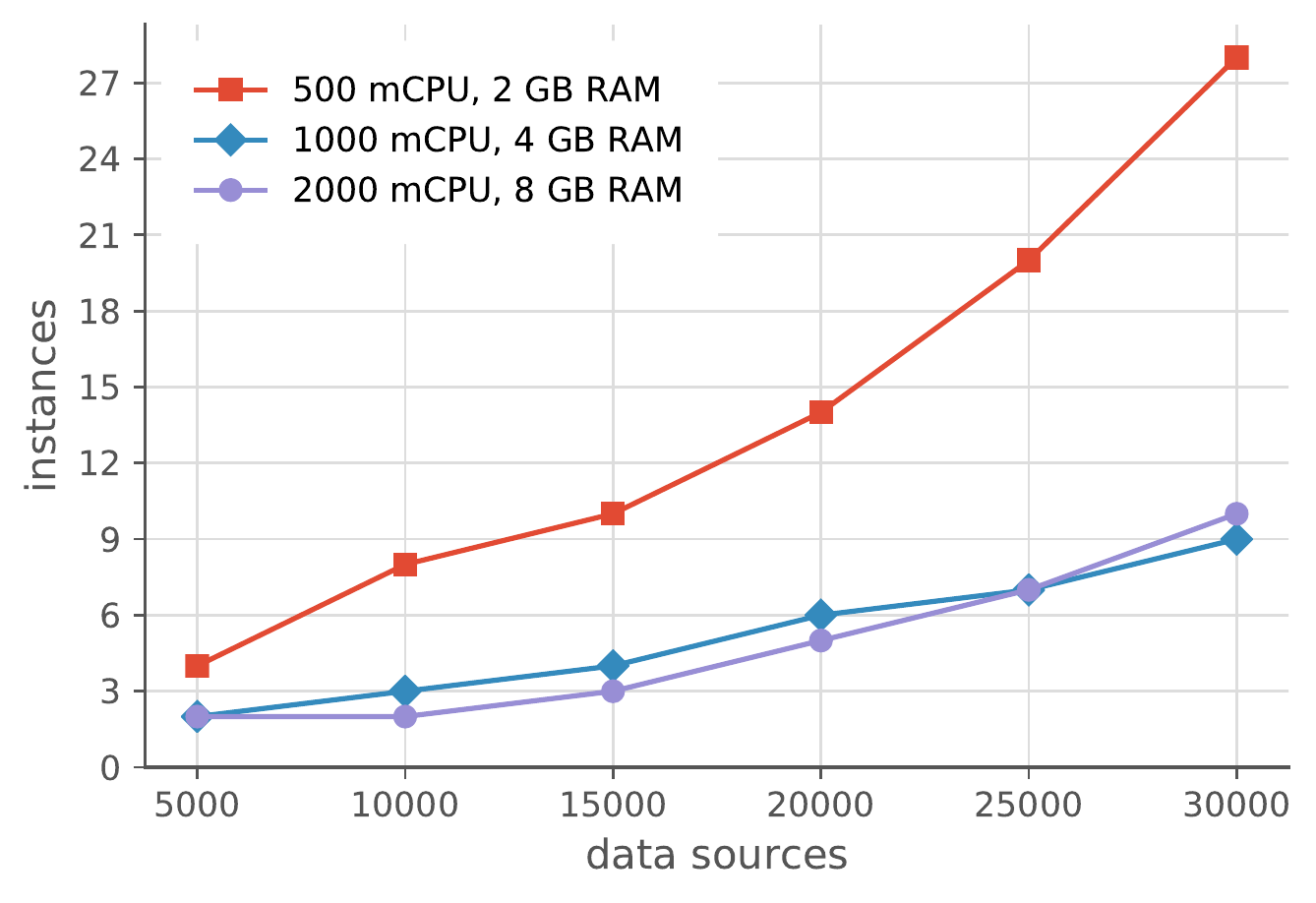}%
		\caption{Use case UC3: Aggregating Time Attributes}
	\end{subfigure}	
	\hfill
	\begin{subfigure}[b]{0.49\textwidth}
		\centering
		\includegraphics[width=\textwidth]{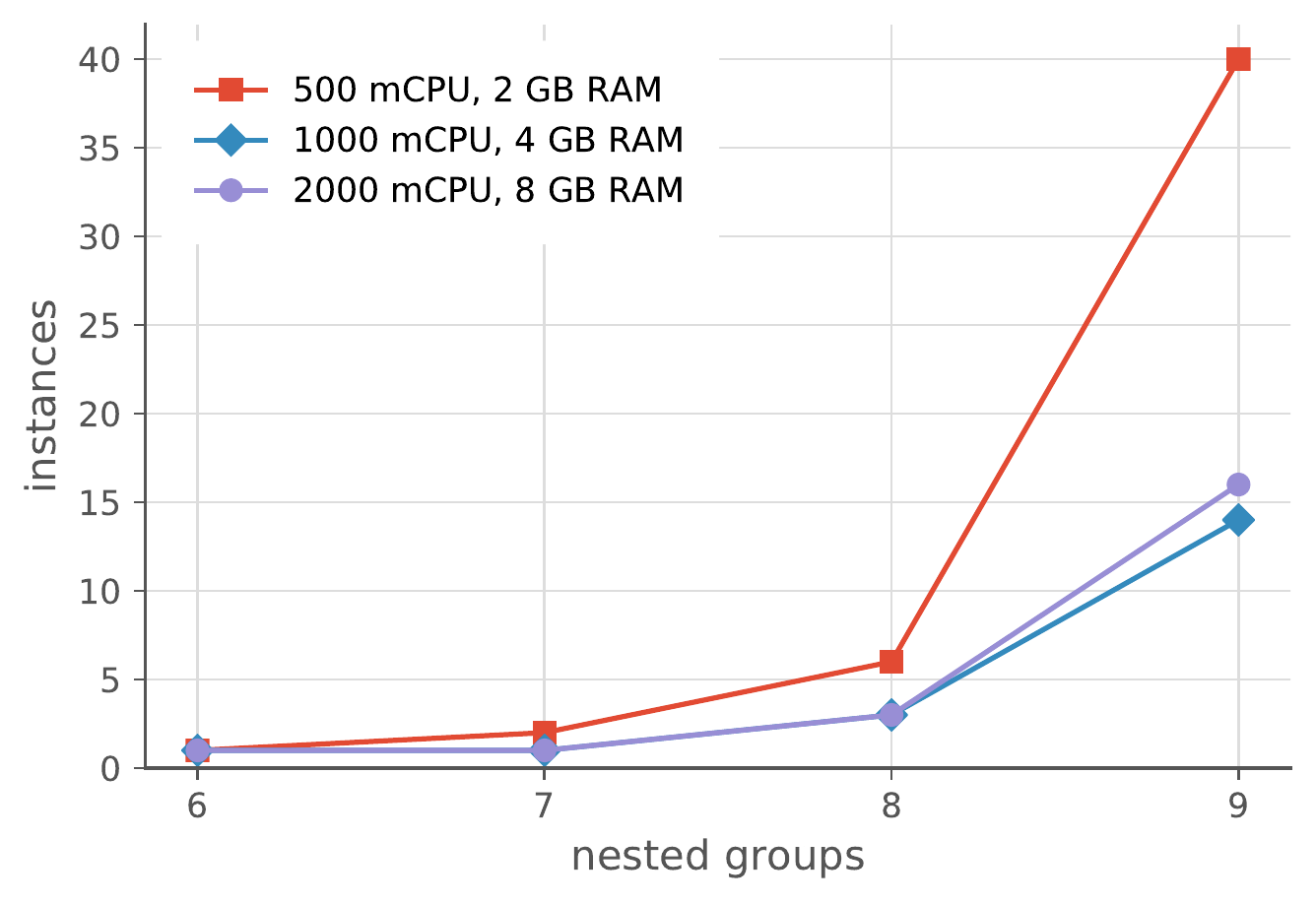}%
		\caption{Use case UC4: Hierarchical Aggregation}
	\end{subfigure}

	\caption{Kafka Streams scalability benchmark results for different computing resources provided for Kafka Streams instances.}
	\label{fig:eval-resources}
\end{figure*}

Figure~\ref{fig:eval-resources} shows our experiment results for the different resource restrictions we applied. We observe that Kafka Streams scales independently of the available resources.  
If using instances restricted to 0.5~cores, significantly more instances are required to process the same workloads compared to the restriction to 1~core.
Use case UC1 requires approximately double the amount of instances. For the other use cases, more than double the amount of instances are required. For use case UC2, the resource demand increases significantly stronger than linearly.
A larger amount of 0.5~core instances allows for more fine-grained scaling (e.g., elastically at runtime) and better fault-tolerance (when distributed among multiple computing nodes). However, these advantages have to be weighed against the introduced overhead. In public clouds, also a cost parameter has to be introduced \cite{Papadopoulos2019}.

Whether restricting the CPU usage to 1 or 2~cores has barely any influence on the resource demand. Even though Kafka Streams runs more threads in addition to the actual processing thread, this overhead seems to be negligible. Thus, we assume the deployment option of using 1~core and 4\,GB memory to be more suitable in most cases.

\subsubsection{Flink Checkpointing}

Checkpointing is Apache Flink's mechanism to ensure fault tolerance \cite{Carbone2017}. With checkpointing enabled, Flink periodically stores the state of all operators as well as the current processing position in data streams.
Furthermore, when using Kafka as a data source, this position in input data streams is also written back to Kafka. Similar to our benchmark of Kafka Streams' commit interval, we benchmark how scalability is effected by using a 10~second checkpointing interval and by disabling checkpoints.

\begin{figure*}[tb]
	\begin{subfigure}[b]{0.49\textwidth}
		\centering
		\includegraphics[width=\textwidth]{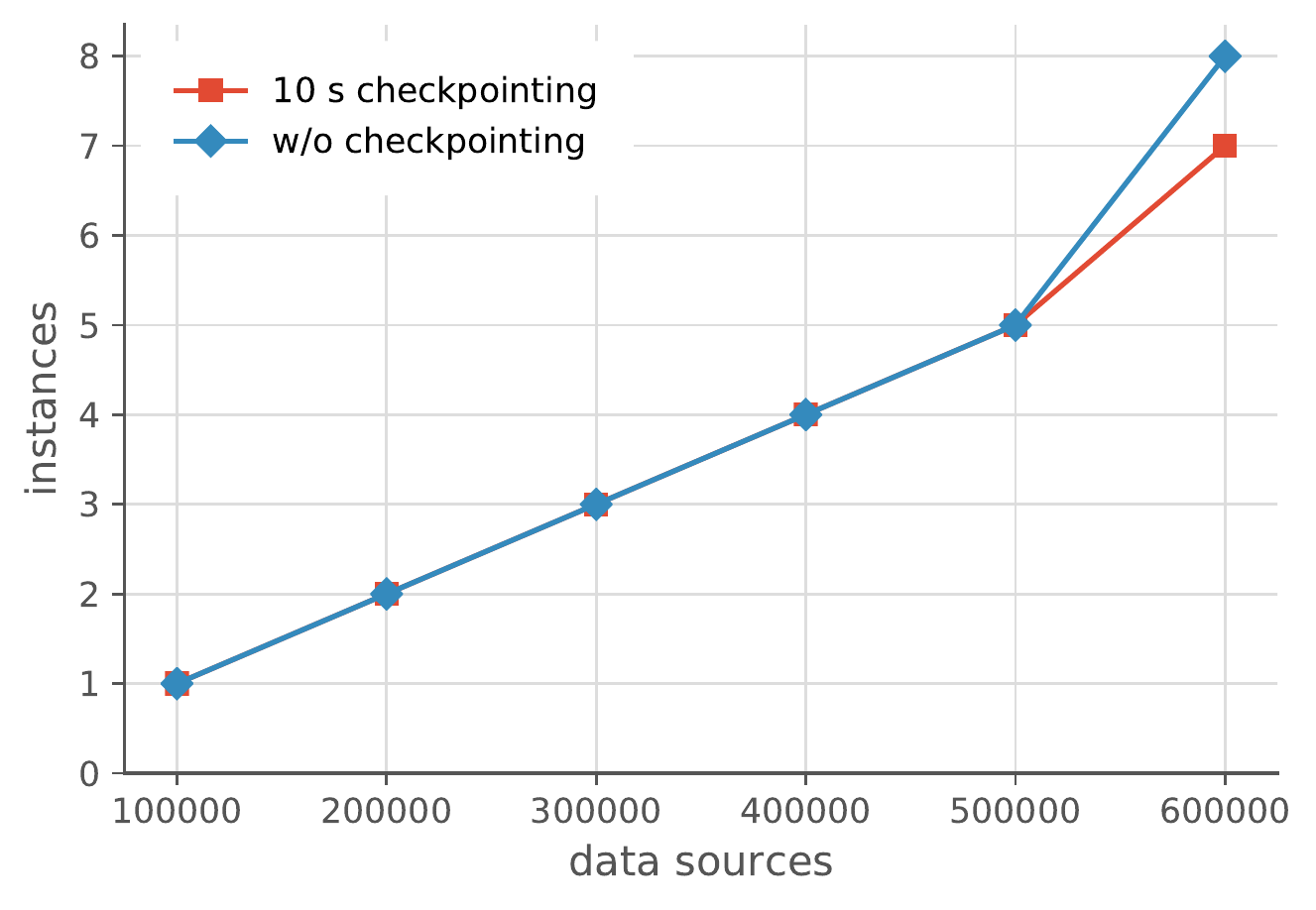}%
		\caption{Use case UC1: Database Storage}
	\end{subfigure}
	\hfill
	\begin{subfigure}[b]{0.49\textwidth}
		\centering
		\includegraphics[width=\textwidth]{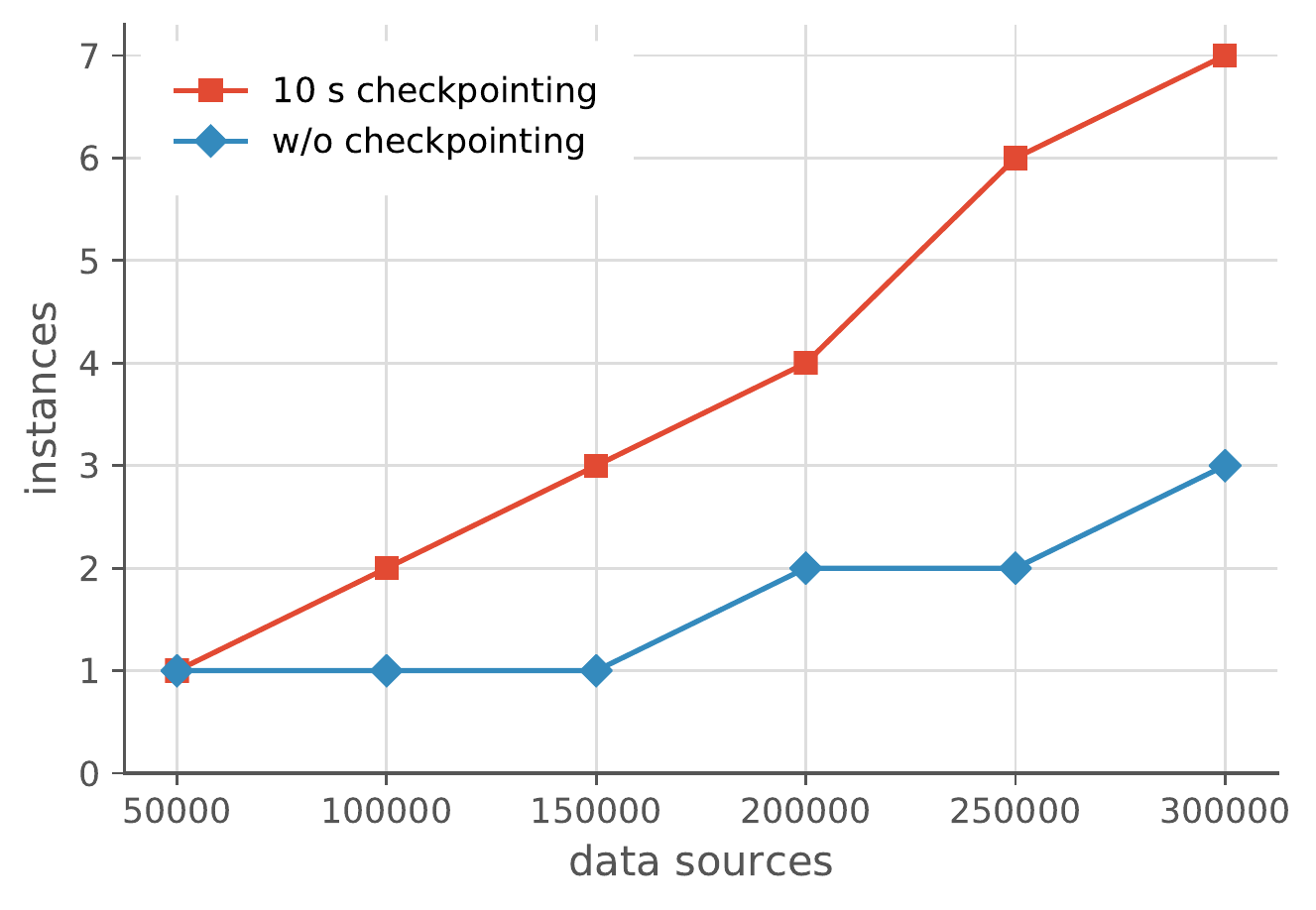}%
		\caption{Use case UC2: Downsampling}
	\end{subfigure}
	
	\vspace{1em}
	
	\begin{subfigure}[b]{0.49\textwidth}
		\centering
		\includegraphics[width=\textwidth]{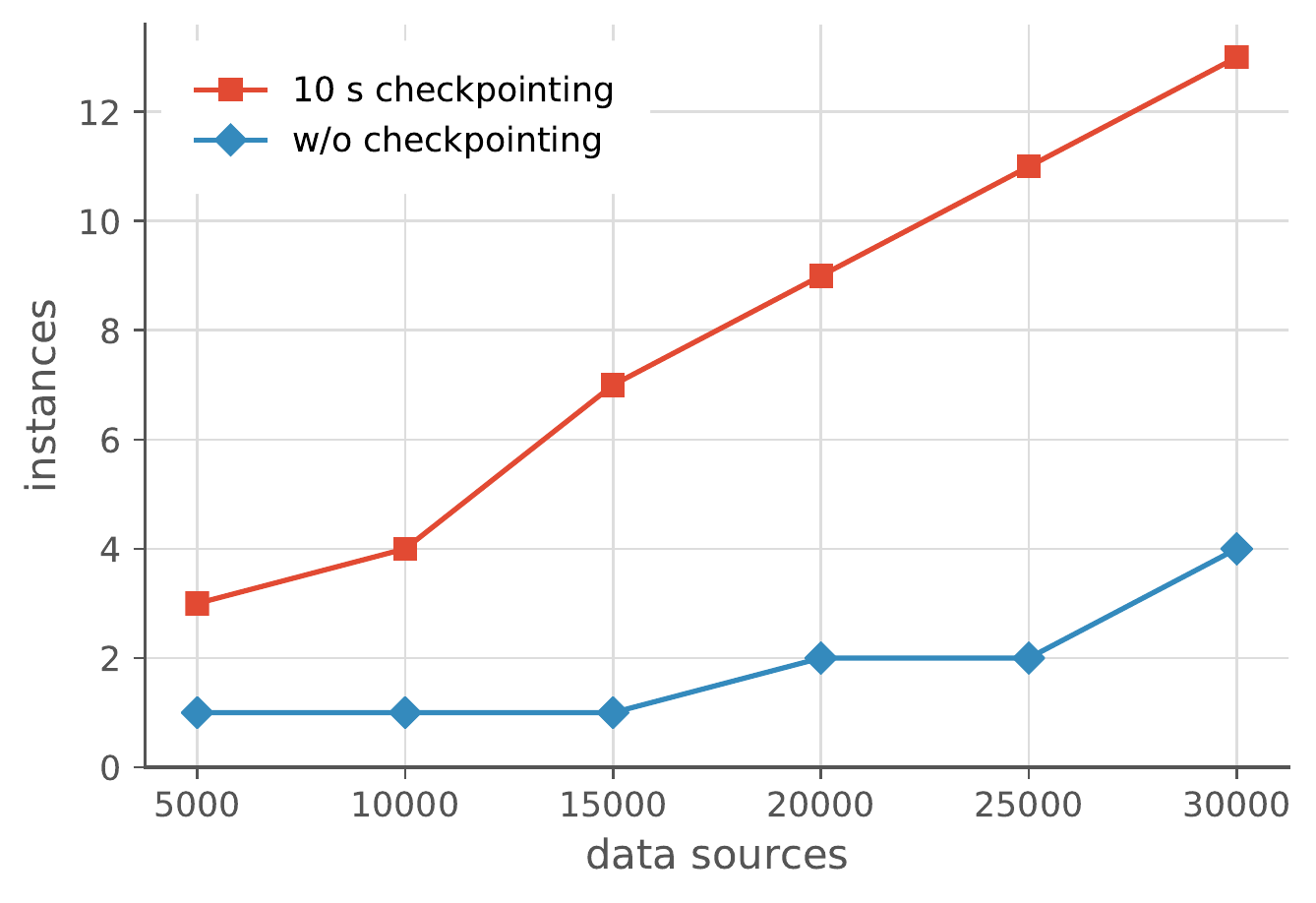}%
		\caption{Use case UC3: Aggregating Time Attributes}
	\end{subfigure}
	\hfill
	\begin{subfigure}[b]{0.49\textwidth}
		\centering
		\includegraphics[width=\textwidth]{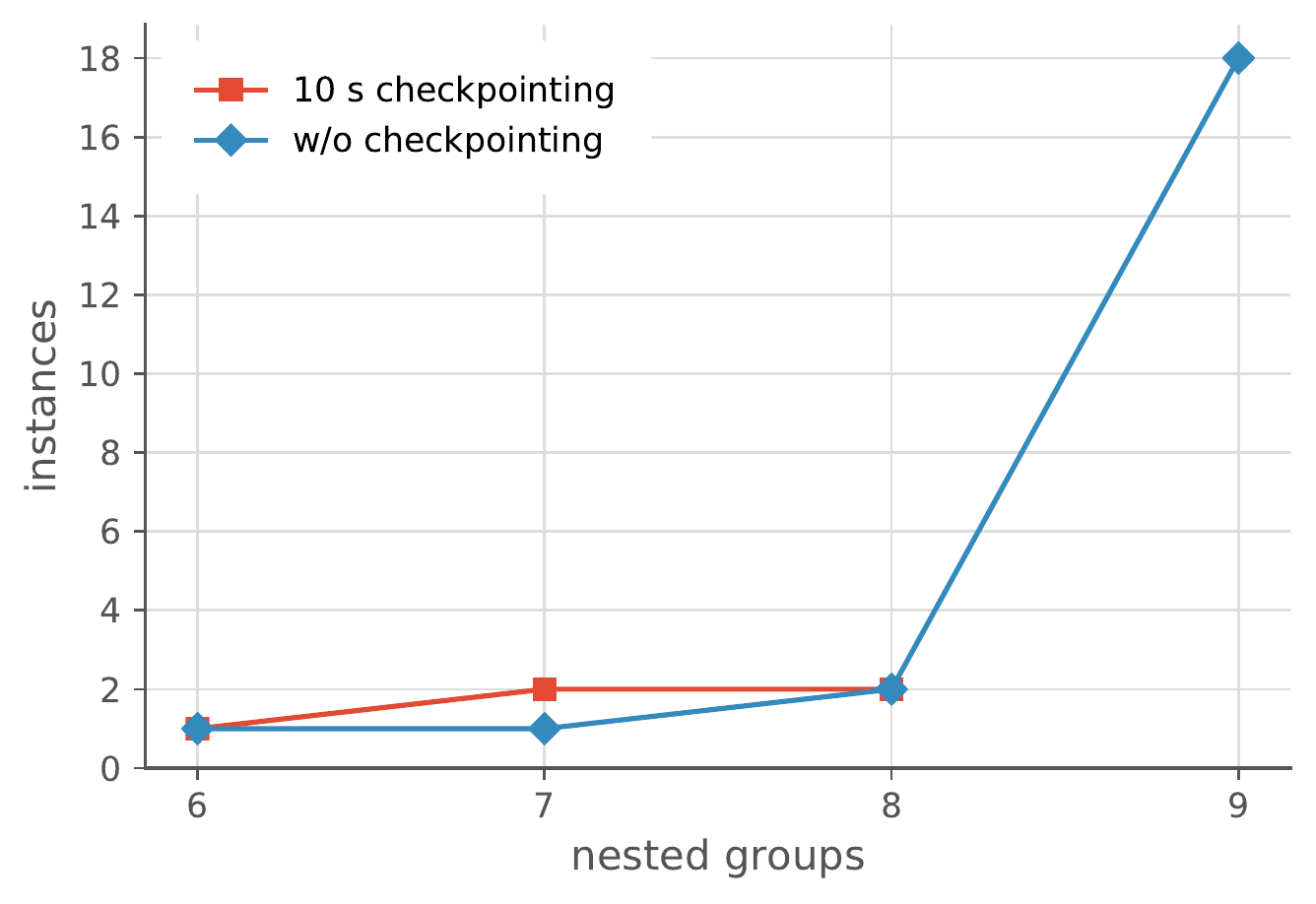}%
		\caption{Use case UC4: Hierarchical Aggregation}
	\end{subfigure}
	\caption{Flink scalability benchmark results for different Flink checkpointing configurations.}
	\label{fig:eval-checkpointing}
\end{figure*}

Figure~\ref{fig:eval-checkpointing} shows the results of our experiments with different checkpointing configurations.
The resource demand of use case UC1, which is stateless, is almost identical, regardless of enabling checkpointing.
Also for the other use cases, resource demand increases linearly with increasing load for both checkpointing enabled and disabled. However, for these use cases, the resource demand with checkpointing enabled is significantly higher than with checkpointing disabled. We were not able to find a sufficient number of instances for use case UC4 with checkpointing enabled when generating records for 9 nested groups, which corresponds to approximately $260\,000$ records per second. Apart from that, we conclude that Flink's fault tolerance mechanism comes to the costs of increased resource usage, but does not restrict scalability.

\subsubsection{Comparison of Kafka Streams and Flink}

Finally, we compare the scalability results of Kafka Streams and Flink. To make the number of required instances comparable for both stream processing engines, we use a 100\,ms commit interval for Kafka Streams and a 100\,ms checkpointing interval for Flink. Note that while Kafka Streams' commit interval controls both latency and fault tolerance, Flink's checkpointing only serves for fault tolerance. For production deployments, less frequent checkpointing than configured in this evaluation is therefore more likely. However, as can be seen in comparison to the previous Flink experiments, there is little difference between using a 100\,ms or a 10\,s checkpointing interval.

\begin{figure*}[tb]
	\begin{subfigure}[b]{0.49\textwidth}
		\centering
		\includegraphics[width=\textwidth]{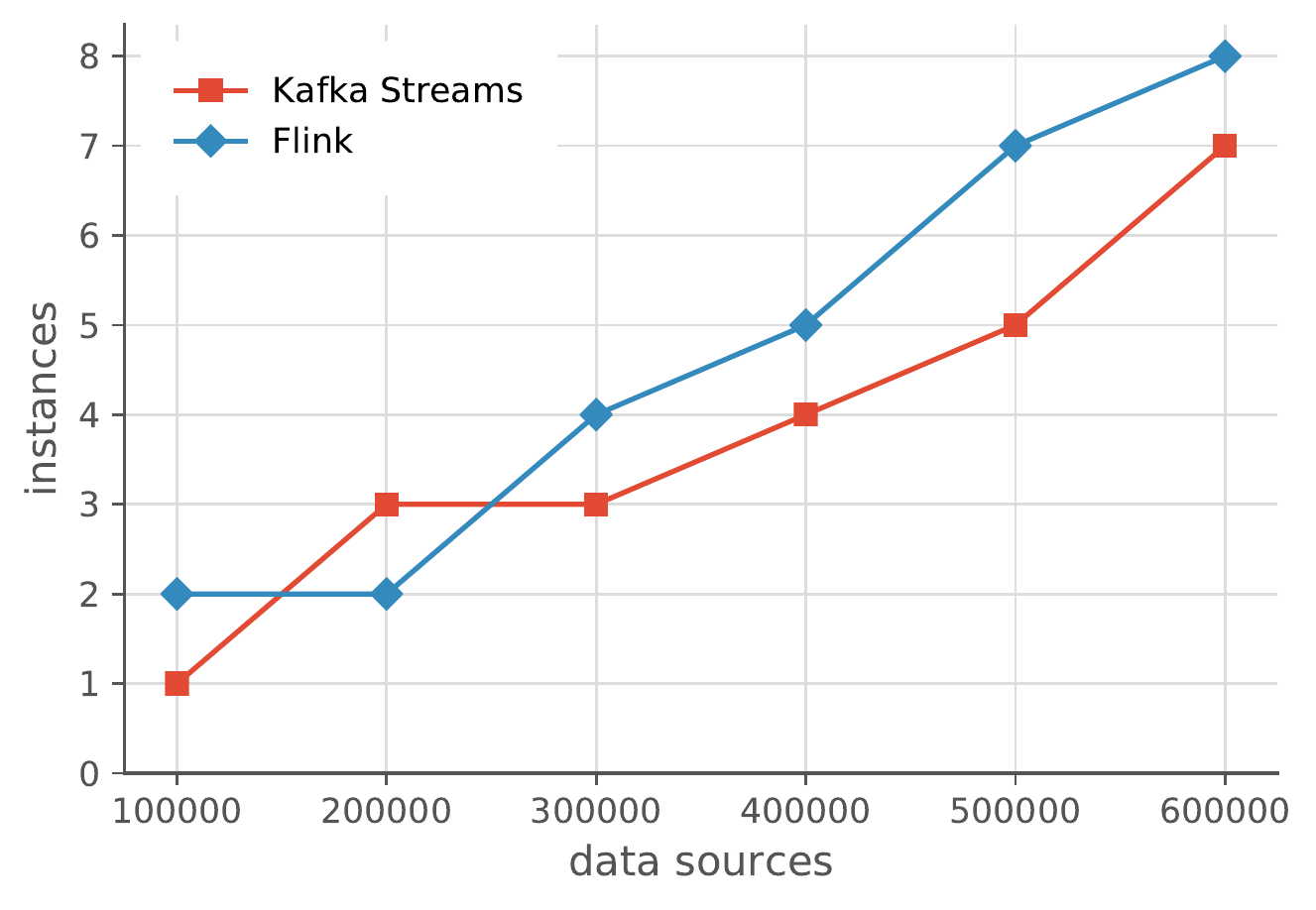}%
		\caption{Use case UC1: Database Storage}
	\end{subfigure}
	\hfill
	\begin{subfigure}[b]{0.49\textwidth}
		\centering
		\includegraphics[width=\textwidth]{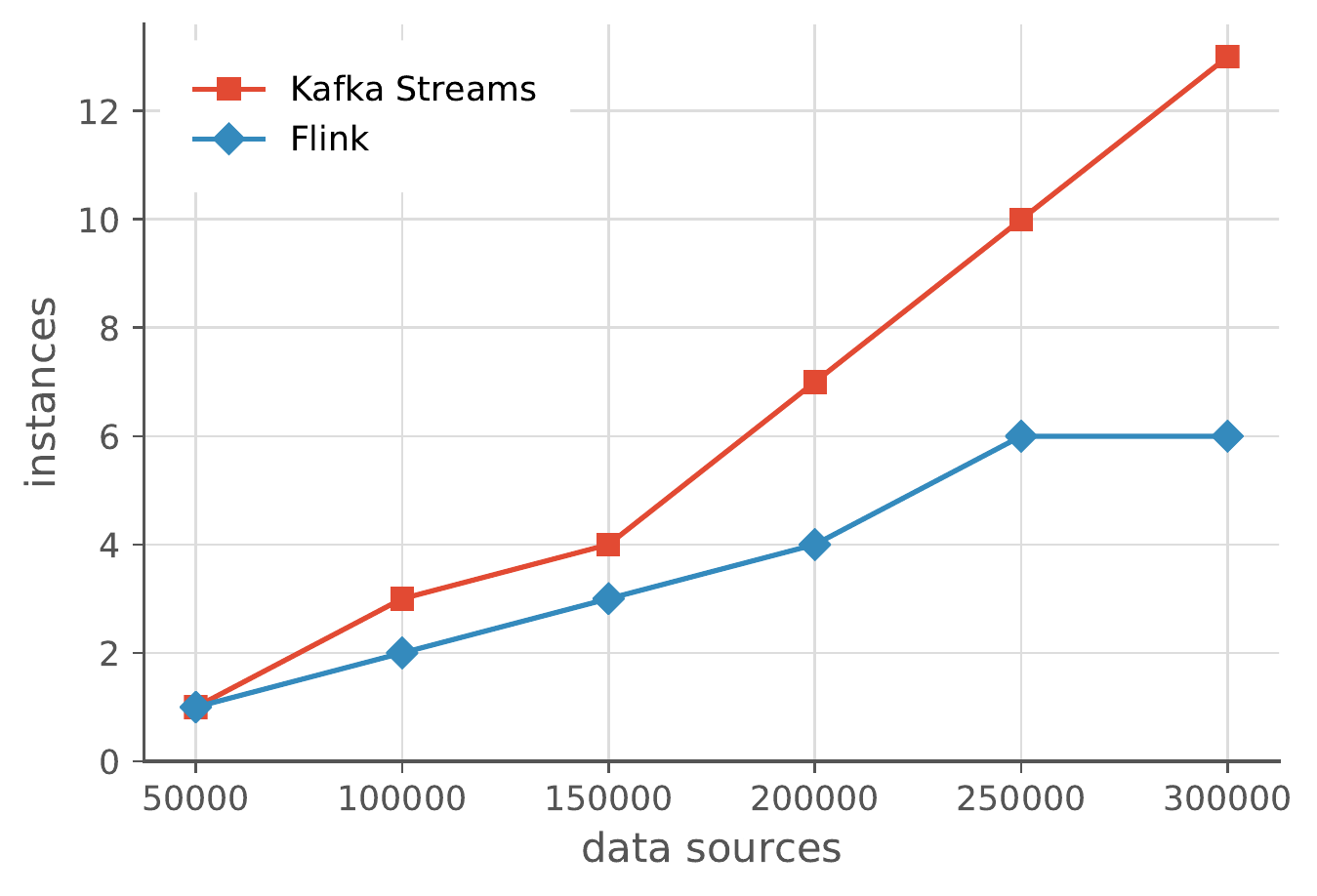}%
		\caption{Use case UC2: Downsampling}
	\end{subfigure}
	
	\vspace{1em}
	
	\begin{subfigure}[b]{0.49\textwidth}
		\centering
		\includegraphics[width=\textwidth]{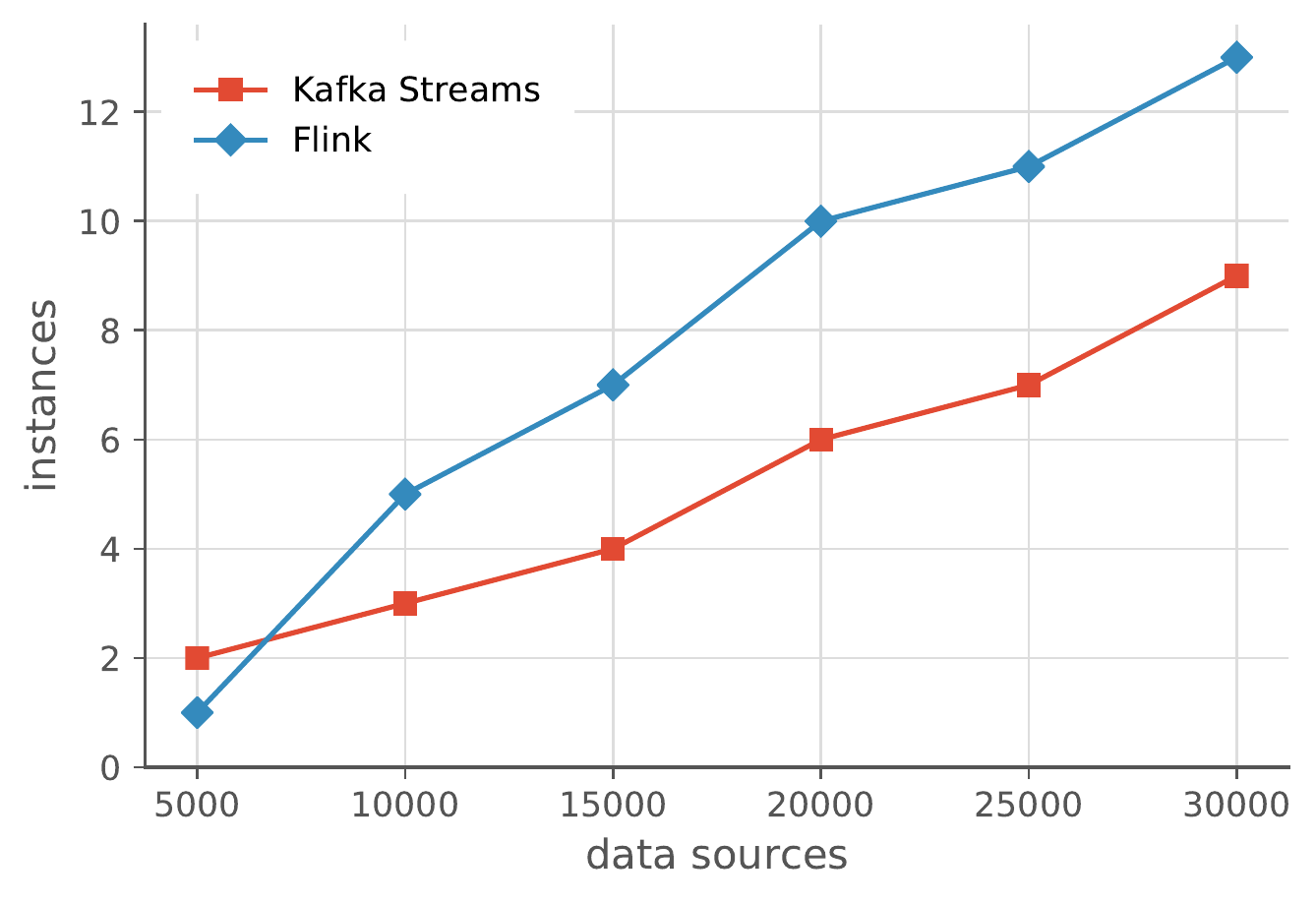}%
		\caption{Use case UC3: Aggregating Time Attributes}
	\end{subfigure}
	\hfill
	\begin{subfigure}[b]{0.49\textwidth}
		\centering
		\includegraphics[width=\textwidth]{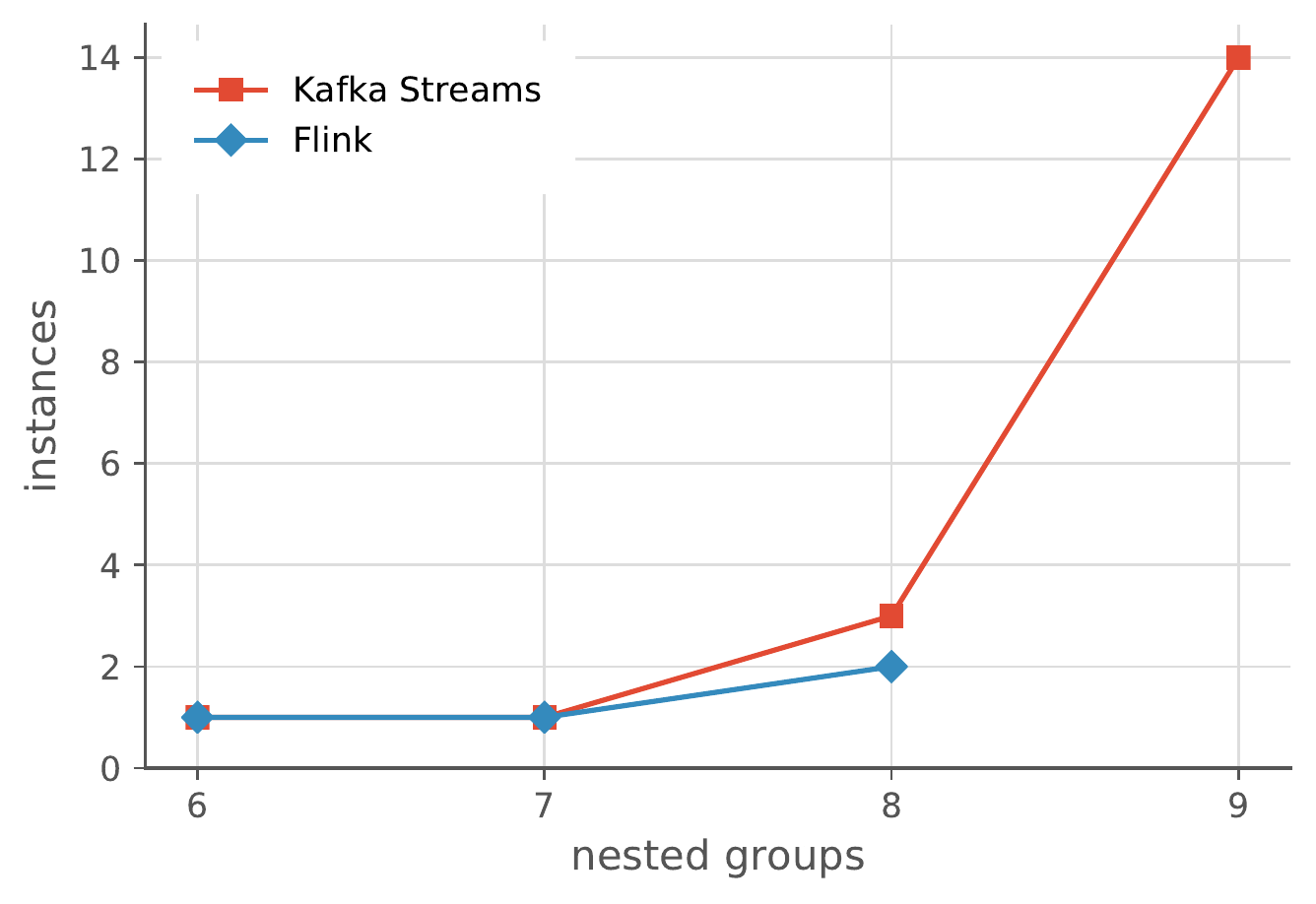}%
		\caption{Use case UC4: Hierarchical Aggregation}
	\end{subfigure}
	\caption{Comparison of scalability benchmark results for Kafka Streams and Flink with 100\,ms commit or checkpointing interval, respectively.}
	\label{fig:eval-comparison}
\end{figure*}

Figure~\ref{fig:eval-comparison} compares the experiment results for Kafka Streams and Flink. As already shown in the previous experiments, both stream processing engines scale approximately linearly. For use cases UC1 and UC3, Flink requires constantly slightly more instances than Kafka Streams, whereas for use case UC2 the resource demand of Kafka Streams is higher. As in the previous evaluation of checkpointing configurations, we were not able to find a sufficient number of instances for use case UC4 with Flink.
In summary, scalability of both stream processing engines is quite similar. 
As the differences are only marginal, our experimental results do not yet allow to rank the scalability of Flink and Kafka Streams (see threats to validity in following section).
However, they show that Theodolite is able to assess the scalability and can be used for more comprehensive comparisons of stream processing engines.

\subsection{Threats to Validity}\label{sec:threats-to-validity}

The primary goal of the experiments in this section is to evaluate whether our proposed method including our identified use cases can be applied for benchmarking scalability. This section aims not for providing an in-depth scalability analysis of Kafka Streams or Flink.
We only conduct our experiments in one, private cloud, which also provides only computing nodes of one hardware configuration. In particular for large amounts of instances, we cannot rule out that scalability is affected by the available hardware. In order to provide more general statements, our experiments should additionally be repeated in other cloud environments \cite{Papadopoulos2019}.
We provide a replication package \cite{ReplicationPackage} as well as all our implementations as open source to simplify replication.
To further increase their validity, experiments should also be repeated several times \cite{Papadopoulos2019}, for example, to rule out the influence of Kubernetes' assignment of pods to nodes.

\section{Conclusions and Future Work}\label{sec:Conclusions}

In this paper, we present the Theodolite method for benchmarking the scalability of distributed stream processing engines. With our method, individual benchmarks are designed based on use cases for stream processing within microservices. Further, our method supports evaluating scalability independently along different dimensions of increasing workloads. 

We propose benchmarks for 4 different use cases of stream processing and 7 different workload dimensions.
We provide implementations for 4 benchmarks with Kafka Streams and Apache Flink as well as a ready-to-use cloud-native implementation of our benchmarking framework.
Our experimental evaluation demonstrates that our benchmarking method is able to assess how a stream processing engine scales with increasing workloads. In particular, it shows that our selected benchmarks cover use cases of different complexity.
Further, our experimental results show that Kafka Streams and Apache Flink can be considered scalable for all evaluated use cases and deployment options. However, we observe that the choice of deployment options has a huge impact on the degree it scales with.

With this paper, we lay the foundation for conducting comprehensive scalability evaluations of stream processing engines.
Besides benchmarking different deployment options, our proposed method can also be used to compare different stream processing engines and to assess scalability along different workload dimensions.
Identifying more complex stream processing use cases such as forecasting or anomaly detection can serve for designing benchmarks, which provide further scalability insights.
For stream processing in edge or fog computing \cite{Hiessl2019}, scalability can also be benchmarked at different levels.
Combined with methods from search-based software engineering, we plan to use our benchmarking method to automate tuning stream processing engine configurations for scalability.
A major prerequisite for this is to execute benchmarks time-efficiently. We plan to supplement our exact measurement method by a heuristic that evaluates results already at runtime to skip unnecessary subexperiments. The experimental results of this paper will serve as reference for the heuristic's quality.
Furthermore, an optimized benchmark execution could be applied to benchmark scalability continuously as part of the DevOps cycle \cite{Waller2015}.


\section*{Acknowledgments}
This research is funded by the German Federal Ministry of Education and Research (BMBF) under grant no.\ 01IS17084 and is part of the Titan project (\url{https://www.industrial-devops.org}). We thank Nico Biernat for implementing the Theodolite benchmarks for Apache Flink.


\bibliography{references}







\end{document}